\documentclass[multphys,vecphys]{svmult}

\usepackage{amsmath}
\usepackage{amsfonts}
\usepackage{amssymb}
\usepackage{bbm}
\usepackage{bbding}
\usepackage{pstcol}
\usepackage{pst-grad}
\usepackage{pst-ghsb}
\usepackage{pstricks}
\usepackage{pst-plot}
\usepackage{fancyhdr}

\usepackage{makeidx}         
\usepackage{graphicx}        
\usepackage{multicol}        
\usepackage[bottom]{footmisc}

\newcommand{\ba}{\begin{array}}
\newcommand{\ea}{\end{array}}

\makeindex             

\begin{document}

\title*{Entanglement, Bell Inequalities\\and Decoherence in Particle
Physics\thanks{Lectures given at \textit{Quantum Coherence in Matter: from Quarks to
Solids}$\,$, 42. Internationale Universit\"atswochen f\"ur Theoretische Physik,
Schladming, Austria, Feb. 28 -- March 6, 2004, Lecture Notes in Physics, Springer Verlag
2005.}}
\author{Reinhold A. Bertlmann\inst{}}
\institute{Institut f\"ur Theoretische Physik, Boltzmanngasse 5, A-1090 Vienna, Austria
\texttt{Reinhold.Bertlmann@univie.ac.at}}
%
%
\maketitle

Abstract\\

We demonstrate the relevance of entanglement, Bell inequalities and decoherence in
particle physics. In particular, we study in detail the features of the ``strange'' $K^0
\bar K^0$ system as an example of entangled meson--antimeson systems. The analogies and
differences to entangled spin--$\frac{1}{2}$ or photon systems are worked out, the
effects of a unitary time evolution of the meson system is demonstrated explicitly. After
an introduction we present several types of Bell inequalities and show a remarkable
connection to $CP$ violation. We investigate the stability of entangled quantum systems
pursuing the question of how possible decoherence might arise due to the interaction of
the system with its ``environment''. The decoherence is strikingly connected to the
entanglement loss of common entanglement measures. Finally, some outlook of the field is
presented.

\section{Introduction}
\label{Introduction}

In 1935, in his famous trilogy on ``The present situation of quantum mechanics'', Erwin
Schr\" odinger \cite{Schrodinger} already realized the peculiar features of what he
called entangled states ---``\ldots\textit{verschr\"ankte Zust\"ande}\ldots'' was
actually his German phrasing--- in connection with quantum systems extended over
physically distant parts. In the same year Einstein, Podolsky and Rosen (EPR) \cite{EPR}
constructed a gedanken experiment for a quantum system of two distant particles to
demonstrate: \textit{Quantum mechanics is incomplete}! Niels Bohr \cite{Bohr} replied
immediately to the EPR article. His message was: \textit{Quantum mechanics is complete}!

Also in 1935, W.H. Furry \cite{Furry} emphasized, inspired by EPR and Schr\" odinger, the
differences between the predictions of quantum mechanics (QM) of non-factorizable systems
and models with spontaneous factorization.

However, the EPR--Bohr debate was regarded as rather philosophical and thus not very
valuable for physicists for about 30 years until John Stewart Bell \cite{bell} brought
this issue up again in his seminal work of 1964 ``On the Einstein--Podolsky--Rosen
Paradox'', which caused a dramatic change in the view about this subject. Bell discovered
what has since become known as \textit{Bell's Theorem:}
\begin{itemize}
    \item [$\bullet$] \textit{No local hidden variable theory can reproduce all
    possible results of QM} !
\end{itemize}

It is achieved by establishing an inequality satisfied by the expectation values of all
\textit{local realistic theories} (LRT) but violated by the predictions of QM.
Inequalities of this type are nowadays named quite generally \textit{Bell inequalities}
(BI).

It is the \textit{nonlocality} arising from quantum entanglement ---the ``\textit{spooky
action at distance}'' as we concluded in the 1980s \cite{Bertlmann}--- which is the basic
feature of quantum physics and is so contrary to our intuition.

Many beautiful experiments have been carried out over the years (see e.g. Refs.
\cite{FreedmanClauser,FryThompson,aspect,WeihsZeilinger}) by using the entanglement of
the polarization of two photons; all (also the long--distance \cite{GisinGroup} or
out-door experiments \cite{AspelmeyerZeilinger}) confirm impressively: \textit{Nature
contains a spooky action at distance}!

The nonlocality does not conflict with Einstein's relativity, so it cannot be used for
superluminal communication, nevertheless, Bell's work \cite{bell,bell2} initiated new
physics, like quantum cryptography \cite{Ekert,DeutschEkert,Hughes,GisinGroup} and
quantum teleportation \cite{Bennett-teleport,ZeilingerTele}, and it triggered a new
technology: quantum information and quantum communication
\cite{ZeilingerInfo1,ZeilingerInfo2}. More about ``from Bell to quantum information'' can
be found in the book \cite{BertlmannZeilinger}.

\subsection{Particle physics}
\label{particlephysics}

Of course, it is of great interest to investigate the EPR--Bell correlations of
measurements also for massive systems in particle physics. Here we want to work out the
analogies and differences to the spin--$\frac{1}{2}$ or photon systems. Already in 1960,
Lee and Yang \cite{lee} and several other authors \cite{Inglis,Day,Lipkin} emphasized the
EPR-like features of the ``strange'' $K^0 \bar K^0$ system in a $J^{PC}=1^{--}$ state,
where the quantum number \textit{strange\-ness} $S=+,-$ plays the role of spin $\Uparrow$
or $\Downarrow$. Indeed many authors
\cite{eberhard,eberhard2,bigi,domenico,Bramon,AncoBramon,BramonGarbarino,Genovese}
suggested to investigate the $K^0 \bar K^0$  pairs which are produced at the $\Phi$
resonance, for instance in the $e^+ e^-$--machine DA$\Phi$NE at Frascati. The
nonseparability of the neutral kaon system ---created in $p\bar{p}$--collisions--- has
been already analyzed by the authors of Refs. \cite{six,CPLEAR-EPR,BGH}.

Similar systems are the entangled beauty mesons, $B^0 \bar B^0$ pairs, produced at the
$\Upsilon$(4S) resonance (see e.g., Refs. \cite{BG1,Dass,BG2,datta,selleribmeson,BG3}),
which we touch only little in this Article.

Specific realistic theories have been constructed
\cite{selleri,SelleriBook,six2,BramonGarbarino2}, which describe the $K^0 \bar K^0$
pairs, as tests versus quantum mechanics. However, a general test of LRT versus QM relies
on Bell inequalities, where ---as we shall see--- the different kaon detection times or
the freely chosen kaon ``quasi--spin'' play the role of the different angles in the
photon or spin--$\frac{1}{2}$ case. Furthermore, an interesting feature of kaons (and
also of B--mesons) is $CP$ violation (charge conjugation and parity) and indeed it turns
out that BI imply bounds on the physical $CP$ violation parameters.

The important difference of the kaon systems as compared to photons is their decay. We
emphasize the necessity of including also the decay product states into the BI in order
to have a unitary time evolution \cite{BertlmannHiesmayr2001}.

A main part of the article is also devoted to the investigation of the stability of the
entangled quantum system. How possible decoherence might arise from the interaction of
the system with its ``environment'', whatever this may be, and we will determine the
strength of such effects. We study how decoherence is related to the loss of
entanglement, and pursue the question: how to detect and quantify entanglement
\cite{BertlmannDurstbergerHiesmayr2002}.

Of course, we cannot cover all subjects of the field. In Section \ref{outlook} we mention
interesting works we could not describe here and give an outlook to what we can expect in
the near future.\\

Thus the contents of our article will be briefly the following:
\begin{itemize}
    \item [$\bullet$] QM of K--mesons
    \item [$\bullet$] Introduction to BI
    \item [$\bullet$] BI for $K$-mesons, in time and ``quasi--spin''
    \item [$\bullet$] Decoherence in entangled $K^0\bar K^0$ system
    \item [$\bullet$] Entanglement measures, entanglement loss and decoherence
\end{itemize}

\section{QM of $K$--mesons}
\label{QM-K-meson}

Neutral K--mesons are wonderful quantum systems! Four phenomena illustrate their
``strange'' behavior that is associated with the work of Abraham Pais
\cite{GellMannPais,PaisPiccioni}:
\begin{enumerate}
    \item[i)] \textit{strangeness}
    \item[ii)] \textit{strangeness oscillation}
    \item[iii)] \textit{regeneration}
    \item[iv)] \textit{$CP$ \textit{violation}}.
\end{enumerate}
Let us start with a discussion of the properties of the neutral kaons, which we need in
the following.

\subsection{Strangeness}
\label{Strangeness}

$K$-mesons are characterized by their \textit{strangeness} quantum number $S$
\begin{eqnarray}
S|K^0\rangle \; &=& + |K^0\rangle \;, \nonumber\\
S|\bar K^0\rangle \; &=& - |\bar K^0\rangle \;.
\end{eqnarray}
As the $K$-mesons are pseudoscalars their parity $P$ is minus and charge conjugation $C$
transforms particle $K^0$ and anti--particle $\bar K^0$ into each other so that we have
for the combined transformation $CP$ (in our choice of phases)
\begin{eqnarray}
CP|K^0\rangle \; &=& - |\bar K^0\rangle \;, \nonumber\\
CP|\bar K^0\rangle \; &=& - |K^0\rangle \;.
\end{eqnarray}
It follows that the orthogonal linear combinations
\begin{eqnarray}\label{K1K2}
|K_1^0\rangle \; &=& \frac{1}{\sqrt{2}}\big\lbrace |K^0\rangle-
|\bar K^0\rangle \big\rbrace \;, \nonumber\\
|K_2^0\rangle \; &=& \frac{1}{\sqrt{2}}\big\lbrace |K^0\rangle+ |\bar K^0\rangle
\big\rbrace
\end{eqnarray}
are eigenstates of $CP$, a quantum number conserved in strong interactions
\begin{eqnarray}
CP|K_1^0\rangle \; &=& + |K_1^0\rangle \;, \nonumber\\
CP|K_2^0\rangle \; &=& - |K_2^0\rangle \;.
\end{eqnarray}

\subsection{$CP$ \textit{violation}}
\label{CPviolation}

Due to weak interactions, that do not conserve \textit{strangeness} and are in addition
$CP$ \textit{violating}, the kaons decay and the physical states, which differ slightly
in mass, $\Delta m = m_L - m_S = 3.49 \times 10^{-6}$ eV, but immensely in their
lifetimes and decay modes, are the short-- and long--lived states
\begin{eqnarray}\label{kaonSL}
|K_S\rangle \; &=& \frac{1}{N}\big\lbrace p |K^0\rangle-q
|\bar K^0\rangle \big\rbrace \;, \nonumber\\
|K_L\rangle \; &=& \frac{1}{N}\big\lbrace p |K^0\rangle+q |\bar K^0\rangle \big\rbrace
\;.
\end{eqnarray}
The weights $p=1+\varepsilon$, $\,q=1-\varepsilon,\,$ with $N^2=|p|^2+|q|^2$ contain the
complex $CP$ \textit{violating parameter} $\varepsilon$ with
$\lvert\varepsilon\rvert\approx10^{-3}$. $CPT$ \textit{invariance} is assumed; thus the
short-- and long--lived states contain the same $CP$ violating parameter
$\varepsilon_S=\varepsilon_L=\varepsilon$. Then the $CPT$ Theorem
\cite{Bell-CPT,Pauli,Luders} implies that time reversal $T$ is violated too.

The short--lived K--meson decays dominantly into $K_S\longrightarrow 2 \pi$ with a width
or lifetime $\Gamma^{-1}_S\sim\tau_S = 0.89 \times 10^{-10}$ s and the long--lived
K--meson decays dominantly into $K_L\longrightarrow 3 \pi$ with $\Gamma^{-1}_L\sim\tau_L
= 5.17 \times 10^{-8}$ s. However, due to $CP$ violation we observe a small amount
$K_L\longrightarrow 2 \pi$. To appreciate the importance of $CP$ violation let us remind
that the enormous disproportion of matter and antimatter in our universe is regarded as a
consequence of $CP$ violation that occurred immediately after the Big Bang.

\subsection{Strangeness oscillation}
\label{StrangenessOscillation}

$K_S, K_L$ are eigenstates of the non--Hermitian ``effective mass'' Hamiltonian
\begin{equation}\label{hamiltonian}
H \, = \, M - \frac{i}{2} \,\Gamma
\end{equation}
satisfying
\begin{equation}
H \,|K_{S,L}\rangle \; = \; \lambda_{S,L} \,|K_{S,L}\rangle
\end{equation}
with
\begin{equation}
\lambda_{S,L} \, = \, m_{S,L} - \frac{i}{2} \,\Gamma_{S,L} \;.
\end{equation}

Both mesons $K^0$ and $\bar K^0$ have transitions to common states (due to $CP$
violation) therefore they mix, that means they \textit{oscillate} between $K^0$ and $\bar
K^0$ before decaying. Since the decaying states evolve ---according to the
Wigner--Weisskopf approximation--- exponentially in time
\begin{equation}
| K_{S,L} (t)\rangle \; = \; e^{-i \lambda_{S,L} t} | K_{S,L} \rangle \;,
\end{equation}
the subsequent time evolution for $K^0$ and $\bar K^0$ is given by
\begin{eqnarray}
| K^0(t) \rangle  &=& g_{+}(t) | K^0 \rangle  + \frac{q}{p} g_{-}(t) | \bar K^0 \rangle
\;, \nonumber\\
| \bar K^0(t) \rangle  &=& \frac{p}{q} g_{-}(t) | K^0 \rangle + g_{+}(t) | \bar K^0
\rangle
\end{eqnarray}
with
\begin{equation}\label{g+-}
g_{\pm}(t) \, = \, \frac{1}{2} \left[ \pm e^{-i \lambda_S t} + e^{-i \lambda_L t} \right]
\;.
\end{equation}
Supposing that a $K^0$ beam is produced at $t=0$, e.g. by strong interactions, then the
probability for finding a $K^0$ or $\bar K^0$ in the beam is calculated to be
\begin{eqnarray}
\left| \langle K^0 | K^0(t) \rangle \right|^2 &=& \frac{1}{4} \big\lbrace e^{-\Gamma_S t}
+ e^{-\Gamma_L t} + 2 \, e^{-\Gamma t}
\cos(\Delta m t)\big\rbrace \;, \nonumber\\
\left| \langle \bar K^0 | K^0(t) \rangle \right|^2 &=& \frac{1}{4} \frac{|q|^2}{|p|^2}
\big\lbrace e^{-\Gamma_S t} + e^{-\Gamma_L t} - 2 \, e^{-\Gamma t} \cos(\Delta m
t)\big\rbrace \, ,
\end{eqnarray}
with $\Delta m=m_L-m_S\,$ and $\,\Gamma = \frac{1}{2}(\Gamma_L+\Gamma_S)\,$.

The $K^0$ beam oscillates with frequency $\Delta m / 2\pi$, where $\Delta m \, \tau_S =
0.47$. The oscillation is clearly visible at times of the order of a few $\tau_S$, before
all $K_S$ have died out leaving only the $K_L$ in the beam. So in a beam which contains
only $K_0$ mesons at the time $t=0$ the $\bar K_0$ will occur far from the production
source through its presence in the $K_L$ meson with equal probability as the $K_0$ meson.
A similar feature occurs when starting with a $\bar K^0$ beam.

\subsection{Regeneration of $K_S$}
\label{Regeneration}

In a $K$--meson beam, after a few centimeters, only the long--lived kaon state survives.
But suppose we place a thin slab of matter into the $K_L$ beam then the short--lived
state $K_S$ is regenerated because the $K^0$ and $\bar K^0$ components of the beam are
scattered/absorbed differently in the matter.

\section{Analogies and quasi--spin}
\label{SpinAnalogy}

A good \textit{optical analogy} to the phenomenon of strangeness oscillation is the
following situation. Let us take a crystal that absorbs the different polarization states
of a photon differently, say H (horizontal) polarized light strongly but V (vertical)
polarized light only weakly. Then if we shine R (right circular) polarized light through
the crystal, after some distance there is a large probability for finding L (left
circular) polarized light.\\

In comparison with spin--$\frac{1}{2}$ particles, or with photons having the polarization
directions V and H, it is especially useful to work with the ``\textit{quasi--spin}''
picture for kaons \cite{BertlmannHiesmayr2001}, originally introduced by Lee and Wu
\cite{LeeWu} and Lipkin \cite{Lipkin}. The two states $| K^0 \rangle$ and $| \bar K^0
\rangle$ are regarded as the quasi-spin states up $\lvert\Uparrow\rangle$ and down
$\lvert\Downarrow\rangle$ and the operators acting in this quasi--spin space are
expressed by Pauli matrices. So the strangeness operator $S$ can be identified with the
Pauli matrix $\sigma_3$, the $CP$ operator with ($-\sigma_1$) and $CP$ violation is
proportional to $\sigma_2$. In fact, the Hamiltonian (\ref{hamiltonian}) can be written
as
\begin{equation}
H \, = \, a\cdot \mathbf{1} + \vec b \cdot \vec \sigma \;,
\end{equation}
with
\begin{eqnarray}
b_1 = b \cos \alpha, \quad b_2 = b \sin \alpha, \quad b_3 = 0 \;, \nonumber\\
a = \frac{1}{2}(\lambda_L + \lambda_S), \quad b = \frac{1}{2}(\lambda_L - \lambda_S)
\end{eqnarray}
($b_3 = 0$ due to $CPT$ invariance), and the phase $\alpha$ is related to the $CP$
parameter $\varepsilon$ by
\begin{equation}
e^{i\alpha} \, = \, \frac{1-\varepsilon}{1+\varepsilon} \;.
\end{equation}
Summarizing, we have the following K--meson --- spin--$\frac{1}{2}$ --- photon analogy:
\begin{center}
\begin{tabular}{c c c}
  K-meson & \quad\quad spin-$\frac{1}{2}$ & \quad photon \\
  $\lvert K^0\rangle$ & \quad\quad $\lvert\Uparrow\rangle_z$ & \quad $\lvert V\rangle$ \\
  $\lvert\bar K^0\rangle$ & \quad\quad $\lvert\Downarrow\rangle_z$ & \quad $\lvert H\rangle$ \\
  $\lvert K_S\rangle$ & \quad\quad $\lvert\Rightarrow\rangle_y$ & \quad $\lvert L\rangle =
  \frac{1}{\sqrt{2}}(\lvert V\rangle-i\lvert H\rangle)$ \\
  $\lvert K_L\rangle$ & \quad\quad $\lvert\Leftarrow\rangle_y$ & \quad $\lvert R\rangle =
  \frac{1}{\sqrt{2}}(\lvert V\rangle+i\lvert H\rangle)$ \\
\end{tabular}
\end{center}

\textbf{Entangled states:} Quite generally, we call a state \textit{entangled} if it is
not \textit{separable}, i.e. not a convex combination of product states.\\

Now, what we are actually interested in are entangled states of $K^0 \bar K^0$ pairs, in
analogy to the entangled spin up and down pairs, or photon pairs. Such states are
produced by $e^+ e^-$--machines through the reaction $e^+ e^- \to \Phi \to K^0 \bar K^0$,
in particular at DA$\Phi$NE, Frascati, or they are produced in $p\bar p$--collisions,
like, e.g., at LEAR, CERN. There, a $K^0 \bar K^0$ pair is created in a $J^{PC}=1^{--}$
quantum state and thus antisymmetric under $C$ and $P$, and is described at the time
$t=0$ by the entangled state
\begin{eqnarray}\label{entangledK0}
| \psi (t=0) \rangle &=&\frac{1}{\sqrt{2}} \left\{ | K^0 \rangle_l \otimes\! | \bar K^0
\rangle _r - | \bar K^0 \rangle _l \otimes | K^0 \rangle _r \right\}\,,
\end{eqnarray}
which can be rewritten in the $K_S K_L$-basis
\begin{eqnarray}\label{entangledKS-KL}
| \psi (t=0) \rangle&=&
 \frac{N_{SL}}{\sqrt{2}}\left\{ | K_S \rangle_l \otimes\! | K_L \rangle _r -
| K_L \rangle _l \otimes | K_S \rangle _r \right\}\,,
\end{eqnarray}
with $N_{SL}=\frac{N^2}{2pq}$. The neutral kaons fly apart and will be detected on the
left ($l$) and right ($r$) side of the source. Of course, during their propagation the
$K^0 \bar K^0$ oscillate and the $K_S, K_L$ states will decay. This is an important
difference to the case of spin--$\frac{1}{2}$ particles or photons which are quite
stable.

\section{Time evolution --- unitarity}
\label{unitarity}

Now let us discuss more closely the time evolution of the kaon states
\cite{BellSteinberger}. At any instant $t$ the state $| K^0(t) \rangle$ decays to a
specific final state $| f \rangle$ with a probability proportional to the absolute square
of the transition matrix element. Because of the unitarity of the time evolution the norm
of the total state must be conserved. This means that the decrease in the norm of the
state $| K^0(t) \rangle$ must be compensated for by the increase in the norm of the final
states.

So starting at $t=0$ with a $K^0$ meson, the state we have to consider for a complete
$t$-evolution is given by
\begin{eqnarray}
|K^0\rangle\;\stackrel{\longrightarrow}\;\;a(t)|K^0\rangle+b(t)|\bar K^0\rangle+\sum_f
c_f(t) |f\rangle \;, \\
a(t)=g_+(t) \qquad\textrm{and}\qquad b(t)=\frac{q}{p} g_-(t) \;.
\end{eqnarray}
The functions $g_{\pm}(t)$ are defined in Eq.(\ref{g+-}). Denoting the amplitudes of the
decays of the $K^0, \bar K^0 \,$ to a specific final state $f$ by
\begin{eqnarray}
\mathcal{A}(K^0\longrightarrow f)\equiv \mathcal{A}_f\qquad\textrm{and}\qquad
\mathcal{A}(\bar K^0 \longrightarrow f)\equiv \bar{\mathcal{A}}_f \;,
\end{eqnarray}
we have
\begin{eqnarray}
\frac{d}{dt}|c_f(t)|^2=|a(t) \mathcal{A}_f+b(t) \bar{\mathcal{A}}_f|^2 \;,
\end{eqnarray}
and for the probability of the decay $K_0 \to f$ at a certain time $\tau$
\begin{eqnarray}
P_{K^0\longrightarrow f}(\tau)=\int_0^\tau \frac{d}{dt}|c_f(t)|^2 dt \; .
\end{eqnarray}

Since the state $| K^0(t) \rangle$ evolves according to a Schr\"odinger equation with
``effective mass'' Hamiltonian (\ref{hamiltonian}) the decay amplitudes are related to
the $\Gamma$ matrix by
\begin{eqnarray}
\Gamma_{11}=\sum_f |\mathcal{A}_f|^2,\quad \Gamma_{22}=\sum_f
|\bar{\mathcal{A}}_f|^2,\quad \Gamma_{12}=\sum_f \mathcal{A}_f^* \bar{\mathcal{A}}_f \; .
\end{eqnarray}
These are the Bell-Steinberger unitarity relations \cite{BellSteinberger}; they are a
consequence of probability conservation, and play an important
role.\\

For our purpose the following formalism generalized to arbitrary quasi--spin states is
quite convenient \cite{BertlmannHiesmayr2001,ghirardi91}. We describe a complete
evolution of mass eigenstates by a unitary operator $U(t,0)$ whose effect can be written
as
\begin{eqnarray}\label{timeevolution}
U(t,0)\; |K_{S,L}\rangle &=& e^{-i \lambda_{S,L} t}\;|K_{S,L}\rangle +
|\Omega_{S,L}(t)\rangle\;,
\end{eqnarray}
where $|\Omega_{S,L}(t)\rangle$ denotes the state of all decay products. For the
transition amplitudes of the decay product states we then have
\begin{eqnarray}
\langle \Omega_S(t)|\Omega_S(t)\rangle&=& 1-e^{-\Gamma_S t} \;,\\
\langle \Omega_L(t)|\Omega_L(t)\rangle&=& 1-e^{-\Gamma_L t} \;,\\
\langle \Omega_L(t)|\Omega_S(t)\rangle&=&\langle K_L|K_S\rangle
(1-e^{i \Delta m t}e^{-\Gamma t}) \;,\\
\langle K_{S,L}|\Omega_S(t)\rangle&=&\langle K_{S,L}|\Omega_L(t)\rangle=0 \;.
\end{eqnarray}
Mass eigenstates (\ref{kaonSL}) are normalized but due to $CP$ violation not orthogonal
\begin{eqnarray}
\langle K_L|K_S\rangle = \frac{2 Re\{\varepsilon\}}{1+|\varepsilon|^2} =: \delta \;.
\end{eqnarray}

Now we consider entangled states of kaon pairs, and we start at time $t=0$ from the
entangled state given in the $K_S K_L$ basis choice (\ref{entangledKS-KL})
\begin{equation}\label{entangledKS}
|\psi(t=0)\rangle \;=\; \frac{N^2}{2\sqrt{2} p q}\big\lbrace |K_S\rangle_l \otimes
|K_L\rangle_r - |K_L\rangle_l \otimes |K_S\rangle_r \big\rbrace \; .
\end{equation}
Then we get the state at time $t$ from (\ref{entangledKS}) by applying the unitary
operator
\begin{eqnarray}\label{U(t)unitary}
U(t,0) &=& U_l(t,0) \otimes U_r(t,0) \; ,
\end{eqnarray}
where the operators $U_l(t,0)$ and $U_r(t,0)$ act on the subspace of the left and of the
right mesons according to the time evolution (\ref{timeevolution}).
\\

What we are finally interested in are the quantum mechanical probabilities for detecting,
or not detecting, a specific quasi--spin state on the left side $|k_n\rangle_l$ and on
the right side $|k_n\rangle_r$ of the source. For that we need the projection operators
$P_{l,r}(k_n)$ on the left, right quasi--spin states $|k_n\rangle_{l,r}$ together with
the projection operators that act onto the orthogonal states $Q_{l,r}(k_n)$
\begin{eqnarray}
P_l(k_n) \, &=& \, |k_n\rangle_{l\,l} \langle k_n| \qquad \textrm{and} \qquad
P_r(k_n) \;  =  \; |k_n\rangle_{r\,r} \langle k_n| \;,\\
Q_l(k_n) \, &=& \, \mathbf{1} - P_l(k_n) \;\;\quad \textrm{and} \qquad Q_r(k_n) \;  =  \;
\mathbf{1} - P_r(k_n) \;.
\end{eqnarray}
So starting from the initial state (\ref{entangledKS}) the unitary time evolution
(\ref{U(t)unitary}) determines the state at a time $t_r$
\begin{eqnarray}
|\psi(t_r)\rangle &=& U(t_r,0)|\psi(t=0)\rangle \; = \; U_l(t_r,0)\otimes U_r(t_r,0)
|\psi(t=0)\rangle \, .
\end{eqnarray}
If we now measure a certain quasi--spin $k_m$ at $t_r$ on the right side means that we
project onto the state
\begin{eqnarray}
|\tilde{\psi}(t_r)\rangle &=& P_r(k_m) |\psi(t_r)\rangle \;.
\end{eqnarray}
This state, which is now a one--particle state of the left--moving particle, evolves
until $t_l$ when we measure another $k_n$ on the left side and we get
\begin{eqnarray}\label{evolutionexact}
|\tilde{\psi}(t_l, t_r)\rangle &=& P_l(k_n) U_l(t_l,t_r) P_r(k_m) |\psi(t_r)\rangle \;.
\end{eqnarray}
The probability of the joint measurement is given by the squared norm of the state
(\ref{evolutionexact}). It coincides (due to unitarity, composition laws and commutation
properties of $l,r$-operators) with the state
\begin{eqnarray}\label{evolutionfactorized}
|\psi(t_l,t_r)\rangle &=& P_l(k_n) P_r(k_m) U_l(t_l,0) U_r(t_r,0) |\psi(t=0)\rangle \;,
\end{eqnarray}
which corresponds to a factorization of the time into an eigentime $t_l$ on the left side
and into an eigentime $t_r$ on the right side.

Then we calculate the quantum mechanical probability $P_{n,m}(Y, t_l; Y, t_r)$ for
finding a $k_n$ at $t_l$ on the left side {\it and} a $k_m$ at $t_r$ on the right side
and the probability $P_{n,m}(N, t_l; N, t_l)$ for finding $no$ such kaons by the
following norms; and similarly the probability $P_{n,m}(Y, t_l; N, t_r)$ when a $k_n$ at
$t_l$ is detected on the left but $no\, k_m$ at $t_r$ on the right
\begin{eqnarray}
P_{n,m}(Y, t_l; Y, t_r) &=& ||P_l(k_n) P_r(k_m) U_l(t_l,0) U_r(t_r,0)
|\psi(t=0)\rangle||^2 \;,\\
P_{n,m}(N, t_l; N, t_r) &=& ||Q_l(k_n) Q_r(k_m) U_l(t_l,0) U_r(t_r,0)
|\psi(t=0)\rangle||^2 \,,\\
P_{n,m}(Y, t_l; N, t_r) &=& ||P_l(k_n) Q_r(k_m) U_l(t_l,0) U_r(t_r,0)
|\psi(t=0)\rangle||^2 \;.
\end{eqnarray}

\section{Bell inequalities for spin--$\frac{1}{2}$ particles}
\label{Bell-inequalities}

In this section we derive the well-known Bell-inequalities \cite{bell2} and we want to
present the details because of the close analogy between the spin/photon systems and kaon
systems. Let us start with a BI which holds most generally, the CHSH inequality, named
after Clauser, Horne, Shimony and Holt \cite{CHSH}, and then we derive from that
inequality ---with two further assumptions--- the original Bell inequality and the
Wigner--type inequality.

Let $A(n,\lambda)$ and $B(m,\lambda)$ be the definite values of two quantum observables
$A^{QM}(n)$ and $B^{QM}(m)$, measured by \textit{Alice} on one side and by \textit{Bob}
on the other. The parameter $\lambda$ denotes the hidden variables which are not
accessible to an experimenter but carry the additional information needed in a LRT. The
measurement result of one observable is $A(n,\lambda) = \pm 1$ corresponding to the spin
measurement ``spin up'' and ``spin down'' along the quantization direction $n$ of
particle $1$; and $A(n,\lambda) = 0$ if no particle was detected at all. The analogue
holds for the result $B(m,\lambda)$ of particle $2$.\\

\textbf{Bell's locality hypothesis:} The basic ingredient is the following.
\begin{itemize}
    \item [$\bullet$] \textit{The outcome of Alice's measurement does {\textrm{not}} depend
    on the settings of Bob's instruments; i.e., $A(n,\lambda)$ depends only on the direction
    $n$, but {\textrm{not}} on $m$; and analogously $B(m,\lambda)$ depends only on $m$ but
    \textrm{not} on $n$} !
\end{itemize}

That's the crucial point, for the combined spin measurement we then have the following
expectation value
\begin{eqnarray}\label{averagevalues}
E(n,m)=\int d\lambda\; \;\rho(\lambda) A(n,\lambda) B(m,\lambda) \;,
\end{eqnarray}
with the normalized probability distribution
\begin{eqnarray}\label{norm}
\int d\lambda\; \;\rho(\lambda)=1\, .
\end{eqnarray}
This quantity $E(n,m)$ corresponds to the quantum mechanical expectation value
$E^{QM}(n, m)=\langle A^{QM}(n)\otimes B^{QM}(m)\rangle$.\\

It is straightforward to estimate the absolute value of the difference of two expectation
values (see, for example, Refs. \cite{CHSH,bell3,Clauser}):
\begin{eqnarray}
E(n,m)&-&E(n,m')\; = \int d\lambda \;\rho(\lambda)A(n,\lambda)B(m,\lambda)\,\big\lbrace
1\pm A(n',\lambda)B(m',\lambda)\big\rbrace\nonumber\\
& &-\int d\lambda \;\rho(\lambda)A(n,\lambda)B(m',\lambda)\,\big\lbrace 1\pm
A(n',\lambda) B(m,\lambda)\big\rbrace \;,
\end{eqnarray}
then the absolute value provides
\begin{eqnarray}
|\, E(n,m)-&E(n,m')&| \; \leq \int d\lambda \;\rho(\lambda)\;\big\lbrace 1\pm
A(n',\lambda)
B(m',\lambda)\big\rbrace\nonumber\\
& &\qquad+\int d\lambda \;\rho(\lambda)\,\big\lbrace 1\pm A(n',\lambda)
B(m,\lambda)\big\rbrace \;,
\end{eqnarray}
and with the normalization \eqref{norm} we get
\begin{eqnarray}\label{chsh-inequality-derivation}
|\, E(n,m)-E(n,m')\,|\;\leq 2\;\pm|\, E(n',m')+E(n',m)\,| \;,
\end{eqnarray}
or written more symmetrically
\begin{eqnarray}\label{chsh-inequality}
S \;=\; |\, E(n,m)-E(n,m')\,|+|\, E(n',m')+E(n',m)\,|\; \leq \;2 \;.
\end{eqnarray}
This is the familiar \textit{CHSH-inequality}, derived by Clauser, Horne, Shimony and
Holt \cite{CHSH} in 1969. \textit{Every local realistic hidden variable theory must obey
this inequality}\,!\\

Calculating the quantum mechanical expectation values $E^{QM}(n,m)$ in the spin singlet
state $\lvert\psi\rangle =
\frac{1}{\sqrt{2}}\,\big(\lvert\Uparrow_n\rangle\lvert\Downarrow_m\rangle -
\lvert\Downarrow_n\rangle\lvert\Uparrow_m\rangle\big)$
\begin{eqnarray}\label{expectval-QM}
    E^{QM}(n,m) &=& \langle\psi\rvert A^{QM}(n)\otimes B^{QM}(m)\lvert\psi\rangle\nonumber\\
    &=& \langle\psi\rvert \vec\sigma\cdot\vec n\;\otimes
    \;\vec\sigma\cdot\vec m\lvert\psi\rangle
    = -\cos\phi_{n, m} \;,
\end{eqnarray}
where the $\phi_{n,m}$ are the angles between the two quantization directions $n$ and
$m$, we insert \eqref{expectval-QM} into \eqref{chsh-inequality} and obtain the following
inequality
\begin{eqnarray}\label{chshphoton}
S(n, m, n', m') &=& |\cos\phi_{n,m}-\cos\phi_{n,m'}|\nonumber\\
&+&|\cos\phi_{n',m'}+\cos\phi_{n',m}|\;\leq\;2 \,.
\end{eqnarray}

For certain angles $\phi$ ---the so called \textit{Bell angles}--- inequality
\eqref{chshphoton} is violated! The maximal violation is $2 \sqrt{2}$, achieved by the
Bell angles $\phi_{n,m'}=\frac{3 \pi}{4}$ and
$\phi_{n,m}=\phi_{n',m'}=\phi_{n',m}=\frac{\pi}{4}$.

Experimentally, for entangled photon pairs, inequality \eqref{chshphoton} is violated
under strict Einstein locality conditions in an impressive way, with a result in close
agreement with QM $S_{\textrm{exp}}=2.73\pm0.02$ \cite{WeihsZeilinger}, confirming
previous
experimental results on similar inequalities \cite{FreedmanClauser,FryThompson,aspect}.\\

Now we make two assumptions: perfect correlations $E(n,n)=-1$, no $0$ results, and choose
3 different angles (e.g. $n'=m'$) then inequality \eqref{chsh-inequality-derivation}
gives
\begin{eqnarray}\label{Bell-orginal}
& &| E(n,m)-E(n,n')|\;\leq 2\;\pm\{ \underbrace{E(n',n')}+E(n',m)\}\nonumber\\
& &\hphantom{| E(n,m)-E(n,n')|\;\leq 2\;\pm\{ }\;\;-1\;\forall\; n'\nonumber\\
& &\textrm{or}\nonumber\\
& &| E(n,m)-E(n,n')|\;\leq\; 1+E(n',m).
\end{eqnarray}
This is the famous \textit{Bell's inequality} derived by J.S. Bell in 1964 \cite{bell}.

Finally, we rewrite the expectation value for the measurement of two spin--$\frac{1}{2}$
particles in terms of probabilities $P$
\begin{eqnarray}
E(n,m) &=& P(n \Uparrow;m \Uparrow)+P(n \Downarrow;m \Downarrow) -P(n \Uparrow;m
\Downarrow)-P(n \Downarrow;m \Uparrow)\nonumber\\
&=& -1+4\; P(n \Uparrow;m \Uparrow)\;,
\end{eqnarray}
where we used $P(n \Uparrow;m \Uparrow)=P(n \Downarrow;m \Downarrow),\,P(n \Uparrow;m
\Downarrow)=P(n \Downarrow;m \Uparrow)$ and $\sum P=1$. Then Bell's original inequality
(\ref{Bell-orginal}) turns into \textit{Wigner's inequality}
\begin{eqnarray}\label{Wigner-inequality}
P(n;m)\;\leq\;P(n;n')+P(n';m)\;,
\end{eqnarray}
where the $P$'s are the probabilities for finding the spins up--up on the two sides or
down--down or twisted, up--down and down--up. Note, that the Wigner inequality has been
originally derived by a set--theoretical approach \cite{Wigner}.

\section{Bell inequalities for $K$--mesons}
\label{BI-for-K}

Let us return to the kaon states which we describe within the ``quasi--spin'' picture. We
start again from the state $| \psi (t=0) \rangle$, Eq.(\ref{entangledK0}) or
Eq.(\ref{entangledKS-KL}), of entangled ``quasi--spins'' states and consider its time
evolution $U(t,0) | \psi (0) \rangle$. Then we find the following situation.

\subsection{Analogies and differences}
\label{K-analogies}

When performing two measurements to detect the kaons at the same time at the left side
and at the right side of the source, the probability of finding two mesons with the same
``quasi--spin'' ---i.e. $K^0 K^0$ with strangeness $(+1,+1)$ or $\bar K^0 \bar K^0$ with
strangeness $(-1,-1)$--- is zero.

That means, if we measure at time $t$ a $K^0$ meson on the left side (denoted by $Y$,
yes), we will find with certainty at the same time $t$ $no \, K^0$ on the right side
(denoted by $N$, no). This is an EPR--Bell correlation analogously to the
spin--$\frac{1}{2}$ or photon case, e.g., with polarization V--H (see
Refs.\cite{BertlmannHiesmayr2001,Bramon,GisinGo}).

The analogy would be perfect, if the kaons were stable ($\Gamma _S = \Gamma _L = 0$);
then the quantum probabilities yield the result
\begin{eqnarray}
P(Y,t_l;Y,t_r) &=& P(N,t_l;N,t_r) \; = \; \frac{1}{4}
\big\lbrace 1 - \cos(\Delta m(t_l - t_r))\big\rbrace \;,\nonumber\\
P(Y,t_l;N,t_r) &=& P(N,t_l;Y,t_r) \; = \; \frac{1}{4} \big\lbrace 1 + \cos(\Delta m(t_l -
t_r))\big\rbrace \;.
\end{eqnarray}
It coincides with the probability result of finding simultaneously two entangled
spin--$\frac{1}{2}$ particles in spin directions $\Uparrow$ $\Uparrow$ or $\Uparrow$
$\Downarrow$ along two chosen directions $\vec n$ and $\vec m$
\begin{eqnarray}
P(\vec n,\Uparrow ;\vec m,\Uparrow) &=& P(\vec n,\Downarrow ;\vec m,\Downarrow) \; = \;
\frac{1}{4}
\big\lbrace 1 - \cos \theta \big\rbrace \;,\nonumber\\
P(\vec n,\Uparrow ;\vec m,\Downarrow) &=& P(\vec n,\Downarrow ;\vec m,\Uparrow) \; = \;
\frac{1}{4} \big\lbrace 1 + \cos \theta \big\rbrace \;.
\end{eqnarray}

\textbf{Analogies:} Perfect analogy between times and angles.
\begin{itemize}
    \item [$\bullet$] \textit{The time differences $\Delta m(t_l - t_r)$ in the kaon case
    play the role of the angle differences $\theta$ in the spin--$\frac{1}{2}$ or photon
    case}.
\end{itemize}

\vspace{0.15cm}

\begin{center}
{\footnotesize{\hspace{0.9cm}\textbf{K propagation} \hspace{1.2cm} \textbf{photon propagation}\\
\setlength{\unitlength}{1cm}
\begin{picture}(10.5,1)(-1.75,-0.6)
    \put(1.45,0){\vector(-1,0){0.8}}
    \put(1.75,0){\vector(1,0){0.8}}
    \put(1.6,0){\circle{0.3}}
    \put(0.9,0.1){$K$}
    \put(2.1,0.1){$K$}
    \put(1.50,-0.5){$1^{--}$}
    \put(0.3,-0.5){{\tiny{left}}}
    \put(2.6,-0.5){{\tiny{right}}}
    \put(4.2,-0.5){{\tiny{Alice}}}
    \put(6.4,-0.5){{\tiny{Bob}}}
    \put(5.35,0){\vector(-1,0){0.8}}
    \put(5.65,0){\vector(1,0){0.8}}
    \put(5.5,0){\circle{0.3}}
    \put(5.2,-0.5){singlet}
\end{picture}

{\footnotesize{\hspace{-1.0cm}$\bullet$ $K^0\bar K^0$ oscillation \; $\bullet$ $K_S$,
$K_L$ decay \hspace{1.4cm} $\bullet$ stable }}}}\\
\end{center}

\vspace{0.5cm}

$\bullet$ for $t_l=t_r$: $\;$EPR--like correlation

\setlength{\unitlength}{1cm}
\begin{picture}(10,1)(-1.75,-0.6)
    \put(1.45,0){\vector(-1,0){0.8}}
    \put(1.75,0){\vector(1,0){0.8}}
    \put(1.6,0){\circle{0.3}}
    \put(0.3,0){$K^0$}
    \put(2.7,0){$\bar K^0$}
    \put(1.4,-0.6){NO}
    \put(0.3,-0.6){$K^0$}
    \put(2.7,-0.6){$K^0$}
    \put(5.35,0){\vector(-1,0){0.8}}
    \put(5.65,0){\vector(1,0){0.8}}
    \put(5.5,0){\circle{0.3}}
    \put(4.2,0.1){$\Uparrow\, V$}
    \put(6.6,0.1){$\Downarrow\, H$}
    \put(5.3,-0.6){NO}
    \put(4.2,-0.6){$\Uparrow\, V$}
    \put(6.6,-0.6){$\Uparrow\, V$}
\end{picture}

\vspace{0.5cm}

$\bullet$ for $t_l\neq t_r$: $\;$EPR--Bell correlation

\begin{picture}(10,0.75)(-1.75,-0.2)
    \put(1.45,0){\vector(-1,0){0.8}}
    \put(1.75,0){\vector(1,0){1.3}}
    \put(1.6,0){\circle{0.3}}
    \put(0.3,0.1){$K^0$}
    \put(2.8,0.1){$K^0\, \bar K^0$}
    \put(5.35,0){\vector(-1,0){0.8}}
    \put(5.65,0){\vector(1,0){0.8}}
    \put(5.5,0){\circle{0.3}}
    \put(4.2,0.1){$\Uparrow\, V$}
    \put(6.6,0.1){$\nearrow\, \swarrow\, V\, H$}
\end{picture}

\vspace{0.4cm}

\textbf{Differences:} There are important physical differences.
\begin{enumerate}
\item[i)] While in the spin--$\frac{1}{2}$ or photon case one can test whether
a system is in an arbitrary spin state $\alpha |\Uparrow\rangle + \beta
|\Downarrow\rangle$ one cannot test it for an arbitrary superposition $\alpha |K^0\rangle
+ \beta |\bar K^0\rangle$.

\item[ii)] For entangled spin--$\frac{1}{2}$ particles or photons it is clearly sufficient
to consider the direct product space $H^l_{spin} \otimes H^r_{spin} \,$ to account for
all spin or polarization properties of the entangled system, however, this is not so for
kaons. The unitary time evolution of a kaon state also involves the decay product states
(see Section \ref{unitarity}), therefore one has to include the Hilbert space of the
decay products $H^l_{\Omega} \otimes H^r_{\Omega} \,$ which is orthogonal to the space
$H^l_{kaon} \otimes H^r_{kaon} \,$ of the surviving kaons.
\end{enumerate}

Consequently, the appropriate dichotomic question on the system is: ``\textit{Are you a
$K^0$ or not}$\,$?'' It is clearly different from the question ``\textit{Are you a $K^0$
or a $\bar K^0$}$\,$?'' (as treated, e.g., in Ref.\cite{GisinGo}), since all decay
products
---an additional characteristic of the quantum system--- are ignored by the latter.

\subsection{Bell--CHSH inequality --- general form}
\label{kaon-Bell-CHSH}

Measuring a $\bar K^0$ (it is the antiparticle that is actually measured via strong
interactions in matter) on the left side we can predict with certainty to find at the
same time $no \, \bar K^0$ at the right side. In any LRT this property $no \, \bar K^0$
must be present on the right side irrespective of having the measurement performed or
not. In order to discriminate between QM and LRT we set up a Bell inequality for the kaon
system where now the different times play the role of the different angles in the
spin--$\frac{1}{2}$ or photon case. But, in addition, we also may use the freedom of
choosing a particular quasi--spin state of the kaon, e.g., the strangeness eigenstate,
the mass eigenstate, or the $CP$ eigenstate. Thus an expectation value for the combined
measurement $E(k_n, t_a; k_m, t_b)$ depends on a certain quasi--spin $k_n$ measured on
the left side at a time $t_a$ and on a (possibly different) $k_m$ on the right side at
$t_b$. Taking over the argumentation of Sect. \ref{Bell-inequalities} we derive the
following \textit{Bell--CHSH inequality} \cite{BertlmannHiesmayr2001}
\begin{eqnarray}\label{chsh-inequ-kaon-gen}
|E(k_n, t_a; k_m, t_b) &-& E(k_n, t_a; k_{m'}, t_{b'})|\nonumber\\
&+& |E(k_{n'}, t_{a'}; k_{m'}, t_{b'}) + E(k_{n'}, t_{a'}; k_m ,t_b)|\; \leq \; 2 \; ,
\end{eqnarray}
which expresses both the freedom of choice in time \textit{and} in quasi--spin. If we
identify $E(k_n, t_a; k_m, t_b)\equiv E(n,m)$ we are back at the inequality
(\ref{chsh-inequality}) for the spin--$\frac{1}{2}$ case.\\

The expectation value for the series of identical measurements can be expressed in terms
of the probabilities, where we denote by $P_{n,m}(Y, t_a; Y, t_b)$ the probability for
finding a $k_n$ at $t_a$ on the left side and finding a $k_m$ at $t_b$ on the right side
and by $P_{n,m}(N, t_a; N, t_b)$ the probability for finding $no$ such kaons; similarly
$P_{n,m}(Y, t_a; N, t_b)$ denotes the case when a $k_n$ at $t_a$ is detected on the left
but $no\; k_m$ at $t_b$ on the right. Then the expectation value is given by the
following probabilities
\begin{eqnarray}\label{expectvalue-prob}
E(k_n, t_a; k_m, t_b)&=& P_{n,m}(Y, t_a; Y, t_b) + P_{n,m}(N, t_a; N, t_b)\nonumber\\
& & - P_{n,m}(Y, t_a; N, t_b) - P_{n,m}(N, t_a; Y, t_b) \, .
\end{eqnarray}
Since the sum of the probabilities for $(Y,Y)$, $(N,N)$, $(Y,N)$ and $(N,Y)$ just add up
to unity we get
\begin{eqnarray}
E(k_n, t_a; k_m, t_b) &=& -1 + 2\, \big\lbrace P_{n,m}(Y, t_a; Y, t_b) + P_{n,m}(N, t_a;
N, t_b)\big\rbrace \, .
\end{eqnarray}
Note that relation (\ref{expectvalue-prob}) between the expectation value and the
probabilities is satisfied for QM and LRT as well.

\subsection{Bell inequality for time variation}
\label{BI-time}

\textbf{Alternative:} In Bell inequalities for meson systems we have an option
\begin{enumerate}
\item[$\bullet$] fixing the quasi--spin --- freedom in time
\item[$\bullet$] freedom in quasi--spin --- fixing time.
\end{enumerate}
Let us elaborate on the first one. We choose a definite quasi--spin, say strangeness
$S=+1$ that means $k_n=k_m=k_{n'}=k_{m'}= K^0$, we neglect $CP$ violation (which does not
play a role to our accuracy level here) then we obtain the following formula for the
expectation value
\begin{eqnarray}\label{Eunitary}
E(t_l;t_r) &=&
-\cos(\Delta m \Delta t)\cdot e^{-\Gamma (t_l+t_r)}\nonumber\\
&&+\frac{1}{2}(1-e^{-\Gamma_L t_l})(1-e^{-\Gamma_S t_r})+\frac{1}{2}(1-e^{-\Gamma_S
t_l})(1-e^{-\Gamma_L t_r}).\hspace{0.3cm}
\end{eqnarray}

Since expectation value (\ref{Eunitary}) corresponds to a unitary time evolution, it
contains, in addition to the pure meson state contribution, terms coming from the decay
product states $|\Omega_{L,S}(t)\rangle$.

However, in the kaon system we can neglect the width of the long--lived $K$--meson as
compared to the short--lived one, $\Gamma_L \ll \Gamma_S$, so that we have to a good
approximation
\begin{equation}\label{Enonunitary}
E^{\textrm{approx}}(t_l;t_r) = -\cos(\Delta m\Delta t)\cdot e^{-\Gamma(t_l+t_r)}\;,
\end{equation}
which coincides with an expectation value where all decay products are ignored (the
probabilities, e.g. $P(K^0,t_a;\bar K^0,t_b)$, just contain the meson states). This is
certainly not the case for other meson systems, like the $B^0 \bar B^0$, $D^0 \bar D^0$
and $B^0_s \bar B^0_s$ systems (see below).

Inserting now the quantum mechanical expectation value (\ref{Enonunitary}) into
inequality (\ref{chsh-inequ-kaon-gen}) we arrive at Ghirardi, Grassi and Weber's result
\cite{ghirardi91}
\begin{eqnarray}\label{chshghirardietal}
& &|e^{-\frac{\Gamma_S}{2} (t_a+t_{a'})} \, \cos(\Delta m (t_a-t_{a'}))
- e^{-\frac{\Gamma_S}{2} (t_a+t_{b'})} \, \cos(\Delta m (t_a-t_{b'}))|\\
& & + |e^{-\frac{\Gamma_S}{2} (t_{a'}+t_b)} \, \cos(\Delta m (t_{a'}-t_b)) +
e^{-\frac{\Gamma_S}{2} (t_b+t_{b'})} \, \cos(\Delta m (t_b-t_{b'}))|\;\leq\;2\,
.\nonumber
\end{eqnarray}
(Of course, we could have chosen $\bar K^0$ instead of $K^0$ without any change).\\

\textbf{No violation:} Unfortunately, inequality (\ref{chshghirardietal}) \textit{cannot
be violated} \cite{ghirardi91,ghirardi92} for any choice of the four (positive) times
$t_a, t_b, t_{a'}, t_{b'}$ due to the interplay between the kaon decay and strangeness
oscillations. As demonstrated in Ref. \cite{trixi} a possible violation depends very much
on the ratio $x = \Delta m /\Gamma\;$. The numerically determined range for \textit{no
violation} is $0<x<2$ \cite{Beatrix-priv} and the experimental value
$x_{\textrm{exper}}=0.95$ lies precisely inside.\\

\textbf{Remark on other meson systems:} Instead of $K$--mesons we also can consider
entangled $B$--mesons, produced via the resonance decay $\Upsilon(4S)\rightarrow B^0\bar
B^0$, e.g., at the KEKB asymmetric $e^+ e^-$ collider in Japan. In such a system, the
\textit{beauty} quantum number $B=+,-$ is the analogue to \textit{strangeness} $S=+,-$
and instead of long-- and short--lived states we have the heavy $|B_H\rangle$ and light
$|B_L\rangle$ as eigenstates of the non--Hermitian ``effective mass'' Hamiltonian. Since
for $B$--mesons the decay widths are equal, $\Gamma_H = \Gamma_L = \Gamma_B$, we get for
the expectation value in a unitary time evolution
\begin{eqnarray}\label{E-B-meson}
E(t_l; ,t_r) \;=\;&&
-\cos(\Delta m_B \Delta t)\cdot e^{-\Gamma_B (t_l+t_r)}\nonumber\\
&&+(1-e^{-\Gamma_B t_l})(1-e^{-\Gamma_B t_r})\;,
\end{eqnarray}
where $\Delta m_B = m_H-m_L$ is the mass difference of the heavy and light B--meson.
Here, the additional term from the decay products cannot be ignored.

Inserting expectation value (\ref{E-B-meson}) into inequality (\ref{chsh-inequ-kaon-gen})
for a fixed quasi--spin, say, for flavor $B=+1$, i.e. $B^0$, we find that the Bell-CHSH
inequality \textit{cannot be violated} in the $x$ range $0<x<2.6$ \cite{Beatrix-priv}.
Again, the experimental value $x_{\textrm{exper}}=0.77$ lies inside.

Precisely the same feature occurs for an other meson--antimeson system, the
\textit{charmed} system $D^0\bar D^0$.

Since the experimental $x$ values for different meson systems are the following ones:
\begin{center}
\begin{tabular}{|r|c|}
\hline
$x\quad\,$ & $\;$ meson system\\
\hline
0.95 & $K^0\bar K^0$ \\
0.77 & $B^0\bar B^0$ \\
$<\;\;$ 0.03 & $D^0\bar D^0$ \\
$>$ 19.00 & $B^0_s \bar B^0_s$ \\
\hline
\end{tabular}
\end{center}\vspace{0.25cm}
\textit{no violation} of the Bell--CHSH inequality occurs for the familiar
meson--antimeson systems; only for the last system a violation is expected.\\

\textbf{Conclusion:} One cannot use the time--variation type of Bell inequality to
exclude
  local realistic theories.

\subsection{Bell inequality for quasi--spin states --- $CP$ violation}
\label{BI-quasispin}

Now we investigate the second option. We fix the time, say at $t=0$, and vary the
quasi--spin of the $K$--meson. It corresponds to a rotation in quasi--spin space
analogously to the spin--$\frac{1}{2}$ or photon case.\\

\textbf{Analogy:} Rotation in $\;$ ``quasi--spin'' space $\quad\longleftrightarrow\quad$
polarization space
\begin{itemize}
  \item [$\bullet$] \textit{The quasi--spin of kaons plays the role of spin or photon
  polarization} !
\end{itemize}
\begin{equation*}
\lvert k\rangle=a\lvert K^0\rangle + b\lvert\bar K^0\rangle \quad\longleftrightarrow\quad
\lvert \vec n\rangle=\cos\frac{\alpha}{2}\lvert \Uparrow\rangle +
\sin\frac{\alpha}{2}\lvert\Downarrow\rangle
\end{equation*}\\

For a BI we need 3 different ``angles'' ---$\;$ ``quasi--spins'' and we may choose the
$H$, $S$ and $CP$ eigenstates
\begin{equation}
|k_n\rangle = |K_S\rangle\,, \qquad |k_m\rangle = |\bar K^0\rangle\,,
\qquad|k_{n'}\rangle = |K_1^0\rangle \;.
\end{equation}
Denoting the probability of measuring the short--lived state $K_S$ on the left side and
the anti--kaon $\bar K^0$ on the right side, at the time $t=0$, by $P(K_S,\bar K^0)$, and
analogously the probabilities $P(K_S,K_1^0)$ and $P(K_1^0,\bar K^0)$ we can easily derive
under the usual hypothesis of Bell's locality the following \textit{Wigner--like Bell
inequality} (see Eq. (\ref{Wigner-inequality}))
\begin{equation}\label{UchiyamaBI}
P(K_S,\bar K^0)\; \leq\; P(K_S,K_1^0) + P(K_1^0,\bar K^0) \; .
\end{equation}
Inequality (\ref{UchiyamaBI}) first considered by Uchiyama \cite{Uchiyama} can be
converted into the inequality $\mbox{Re}\, \{\varepsilon\}\; \leq\; |\, \varepsilon\,|^2
\,$ for the $CP$ violation parameter $\varepsilon \,$, which is obviously violated by the
experimental value of $\varepsilon \,$, having an absolute value of order
$10^{-3}$ and a phase of about $45^\circ$ \cite{ParticleData}.\\

We, however, want to stay as general and loophole--free as possible and demonstrate the
relation of Bell inequalities to $CP$ violation in the following way \cite{BGH-CP}.

The Bell inequality (\ref{UchiyamaBI}) is rather formal because it involves the
unphysical $CP$--even state $| K^0_1 \rangle$, but it implies an inequality on a
\textit{physical} $CP$ violation parameter which is experimentally testable. For the
derivation, recall the $H$ and $CP$ eigenstates, Eqs. \eqref{kaonSL} and \eqref{K1K2},
then we have the following transition amplitudes
\begin{equation}
\langle \bar{K}^0\lvert K_S\rangle=-\frac{q}{N}\,,\;\quad \langle \bar{K}^0\lvert
K_1^0\rangle = - \frac{1}{\sqrt{2}}\,,\;\quad \langle K_S\lvert K_1^0\rangle =
\frac{1}{\sqrt{2}N}(p^*+q^*)\,,
\end{equation}
which we use to calculate the probabilities in BI (\ref{UchiyamaBI}). Optimizing the
inequality we find, independent of any phase conventions of the kaon states,
\begin{eqnarray}\label{inequalpq}
|\,p\,| &\leq& |\,q\,| \;.
\end{eqnarray}

\vspace{0.2cm}

\textbf{Proposition:} $p,q$ -- kaon transition coefficients
\begin{itemize}
  \item [$\bullet$] \textit{Inequality $\,|\,p\,| \leq |\,q\,|\,$ is experimentally
  testable} !\\
\end{itemize}

\textbf{Semileptonic decays:} Let us consider the semileptonic decays of the $K$ mesons.
The strange quark $s$ decays weakly as constituent of $\bar K^0$:

\vspace{0.1cm}

\begin{center}
\includegraphics[height=1cm]{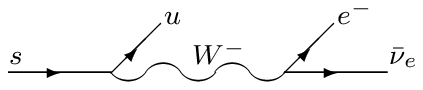}
\end{center}

\vspace{0.3cm}

Due to their quark content the kaon $K^0(\bar s d)$ and the anti--kaon $\bar K^0(s \bar
d)$ have the following definite decays:\\

decay of strange particles\hspace{3.3cm}quark level\\

{\footnotesize{\begin{minipage}[c]{2cm}
\begin{tabular}{cccl}
    & $K^0(d\bar{s})$ & $\longrightarrow$ & $\pi^-(d\bar{u})\quad l^+\;\;\nu_l$ \\
    Q & 0 &&$-1$   \\
    S & 1 & &$\;\,\,0$ \\
\end{tabular}
\end{minipage}\hspace{5cm}
\begin{minipage}[c]{2.5cm}
\begin{tabular}{ccl}
    $\bar{s}$ & $\longrightarrow$ & $\bar{u}\quad l^+\;\;\nu_l$ \\
    $\frac{1}{3}$ &&\hspace{-5pt}$-\frac{2}{3}$   \\
    1 & &$\;0$  \\
\end{tabular}
\end{minipage}}}\\

\vspace{0.2cm}

{\footnotesize{
\begin{minipage}[c]{2cm}
\begin{tabular}{cccl}
    & $\bar{K}^0(\bar{d}s)$ & $\longrightarrow$ & $\pi^+(\bar{d}u)\quad l^-\;\;\bar{\nu}_l$ \\
    Q & $\;\,\,0$ &&$+1$   \\
    S & $-1$ & &$\;\,\,0$  \\
\end{tabular}
\end{minipage}\hspace{4.9cm}
\begin{minipage}[c]{2.5cm}
\begin{tabular}{ccl}
   $s$ & $\longrightarrow$ & $u\quad l^-\;\;\bar{\nu}_l$ \\
   $-\frac{1}{3}$ &&$\frac{2}{3}$  \\
   $-1$ & &\,$0$   \\
\end{tabular}
\end{minipage}}}\\

\rule{4.2cm}{0cm} $\Delta S=\Delta Q \quad $ rule

\vspace{0.4cm}

In particular, we study the leptonic charge asymmetry
\begin{eqnarray}\label{asymlept}
\delta &=& \frac{\Gamma(K_L\rightarrow \pi^- l^+ \nu_l) - \Gamma(K_L\rightarrow \pi^+ l^-
\bar \nu_l)}{\Gamma(K_L\rightarrow \pi^- l^+ \nu_l) + \Gamma(K_L\rightarrow \pi^+ l^-
\bar \nu_l)} \;\quad\textrm{with}\quad\; l=\mu, e \;,
\end{eqnarray}
where $l$ represents either a muon or an electron. The $\Delta S = \Delta Q$ rule for the
decays of the strange particles implies that ---due to their quark content--- the kaon
$K^0(\bar s d)$ and the anti-kaon $\bar K^0(s \bar d)$ have definite decays (see above
Table). Thus, $l^+$ and $l^-$ tag $K^0$ and $\bar K^0$, respectively, in the $K_L$ state,
and the leptonic asymmetry (\ref{asymlept}) is expressed by the probabilities $|p|^2$ and
$|q|^2$ of finding a $K^0$ and a $\bar K^0$, respectively, in the $K_L$ state
\begin{eqnarray}
\delta &=& \frac{|p|^2-|q|^2}{|p|^2+|q|^2} \; .
\end{eqnarray}
Then inequality (\ref{inequalpq}) turns into the bound
\begin{eqnarray}\label{inequaldelta}
\delta &\leq& 0
\end{eqnarray}
for the leptonic charge asymmetry which measures $CP$ violation.

If $CP$ were conserved, we would have $\delta = 0$. Experimentally, however, the
asymmetry is nonvanishing\footnote{It is the weighted average over electron and muon
events, see Ref. \cite{ParticleData}.}, namely
\begin{equation}\label{deltaexp}
\delta = (3.27 \pm 0.12)\cdot 10^{-3} \; ,
\end{equation}
and is thus a clear sign of $CP$ violation.

The bound (\ref{inequaldelta}) dictated by BI (\ref{UchiyamaBI}) is in contradiction to
the experimental value (\ref{deltaexp}) which is definitely positive.
\textit{In this sense $CP$ violation is related to the violation of a Bell inequality} !\\

On the other hand, we can replace $\bar K^0$ by $K^0$ in the BI (\ref{UchiyamaBI}) and
along the same lines as discussed before we obtain the inequality
\begin{eqnarray}\label{inequalqp}
|\,p\,| &\geq& |\,q\,| \qquad \textrm{or} \qquad \delta \geq 0 \;.
\end{eqnarray}
Altogether inequalities (\ref{inequalpq}), (\ref{inequaldelta}) and (\ref{inequalqp})
imply the strict equality
\begin{eqnarray}\label{equalpq}
|\,p\,| &=& |\,q\,| \qquad \textrm{or} \qquad \delta = 0 \;,
\end{eqnarray}
which is in contradiction to experiment.\\

\textbf{Conclusion:}  The premises of LRT are \textit{only} compatible with strict $CP$
conservation in $K^0 \bar K^0$ mixing. Conversely, $CP$ violation in $K^0 \bar K^0$
mixing, no matter which sign the experimental asymmetry (\ref{asymlept}) actually has,
always leads to a \textit{violation} of a BI, either of inequality (\ref{inequalpq}),
(\ref{inequaldelta}) or of (\ref{inequalqp}). In this way, $\delta \neq 0$ is a
manifestation of the entanglement of the considered state.\\

We also want to remark that in case of Bell inequality (\ref{UchiyamaBI}), since it is
considered at $t=0$, it is rather \textit{contextuality} than nonlocality which is
tested. \textit{Noncontextuality} means that the value of an observable does not depend
on the experimental context; the measurement of the observable must yield the value
independent of other simultaneous measurements. The question is whether the properties of
individual parts of a quantum system do have definite or predetermined values before the
act of measurement --- a main hypothesis in hidden variable theories. The \textit{no-go
theorem of Bell--Kochen--Specker} \cite{noncontext} states:
\begin{itemize}
    \item [$\bullet$] \textit{Noncontextual theories are incompatible with QM}$\,$!
\end{itemize}
The contextual quantum feature is verified in our case.

\section{Decoherence in entangled $K^0\bar K^0$ system}
\label{decoherence}

Again, we consider the creation of an entangled kaon state at the $\Phi$ resonance; the
state propagates to the left and right until the kaons are measured.
\vspace{0.2cm}
\setlength{\unitlength}{1cm}
\begin{picture}(20,1.4)(-4,-0.6)
    \put(1.45,0){\vector(-1,0){1.5}}
    \put(1.75,0){\vector(1,0){1.5}}
    \put(1.6,0){\circle{0.3}}
    \put(0.9,0.1){{\footnotesize{$K$}}}
    \put(2.1,0.1){{\footnotesize{$K$}}}
    \put(1.5,-0.6){{\footnotesize{$\Phi$}}}
    \put(-0.05,-0.5){{\footnotesize{left}}}
    \put(2.8,-0.5){{\footnotesize{right}}}
    \put(-0.75,-0.1){\fbox{$k_1$}}
    \put(3.5,-0.1){\fbox{$k_2$}}
\end{picture}
measure\\
quasi-spin: $\qquad\,\lvert k_1\rangle_l\;$ on left side $\;\longleftrightarrow\;$
$\lvert k_2\rangle_r\;$ on right side\\

\vspace{0.3cm}

How can we describe and measure possible decoherence in the entangled state? Decoherence
provides us some information on the quality of the entangled state.

In the following we consider possible decoherence effects arising from some interaction
of the quantum system with its ``environment''. Sources for ``standard'' decoherence
effects are the strong interaction scatterings of kaons with nucleons, the weak
interaction decays and the noise of the experimental setup. ``Nonstandard'' decoherence
effects result from a fundamental modification of QM and can be traced back to the
influence of quantum gravity \cite{Hawking,tHooft1,tHooft2} ---quantum fluctuations in
the space--time structure on the Planck mass scale--- or to dynamical state--reduction
theories \cite{GRW,pearle,gisinP,penrose}, and arise on a different energy scale. We will
present in the following a specific model of decoherence and quantify the strength of
such possible effects with the help of data of existing experiments.

\subsection{Density matrix}
\label{Density matrix}

In decoherence theory there will occur a statistical mixture of quantum states, which can
be elegantly described by a \textit{density matrix}. It is of great importance for
quantum statistics, therefore we briefly recall its conception and basic properties.

Usually a quantum system is described by a state vector $\lvert\psi\rangle$ which is
determined by the Schr\"odinger equation
\begin{equation}\label{Schrodinger}
i\,\frac{\partial}{\partial t}\;\lvert\psi\rangle \;=\; H\,\lvert\psi\rangle\,, \qquad\;
\hbar=1\;.
\end{equation}
The expectation value of an observable $A$ in the state $\lvert\psi\rangle$ is calculated
by
\begin{equation}
\langle A\rangle=\langle\psi\rvert A\lvert\psi\rangle\;.
\end{equation}
Then it is rather suggestive to define a \textit{density matrix for pure states} as
\begin{equation}
\rho=\lvert\psi\rangle\langle\psi\rvert\;,
\end{equation}
with properties
\begin{equation}
\rho^2=\rho\,, \qquad\rho^\dag=\rho\,, \qquad {\rm tr}\rho=1\;.
\end{equation}\\
The extension to a statistical mixture of states with probabilities $p_i$
---the \textit{density matrix for mixed states}--- is straight forward
\begin{equation}
\rho=\sum_i p_i \,\lvert\psi_i\rangle\langle\psi_i\rvert\qquad\mbox{with}\quad p_i\geq0
\quad \sum_i p_i=1\;.
\end{equation}
The \textit{mixed state density matrix} has the properties
\begin{equation}
\rho^2\neq\rho\,, \qquad\rho^\dag=\rho\,, \qquad {\rm tr}\rho=1\,, \qquad {\rm
tr}\rho^2<1\;.
\end{equation}
Then the expectation value of observable $A$ in a state $\rho$ is defined by
\begin{equation}
\langle A\rangle={\rm tr}\,\rho A\;.
\end{equation}
Due to the Schr\"odinger equation (\ref{Schrodinger}) the time evolution of the density
matrix is determined by an equation, called the \textit{von Neumann equation}
\begin{equation}
i\,\frac{\partial}{\partial t}\;\rho \;=\; [H,\rho]\;.
\end{equation}


\textbf{Example:} Density matrix for spin--$\frac{1}{2}$ state
\begin{eqnarray}
\rho = \frac{1}{2}(\mathbbm{1} + \vec\rho\cdot\vec\sigma) \quad &&\textrm{with} \quad
\vec\rho = {\rm tr}\,\rho\,\vec\sigma = \langle\vec\sigma\rangle \quad
\textrm{Bloch vector}\;;\nonumber\\
&&\textrm{if} \quad \lvert\vec\rho\rvert=1 \quad \textrm{pure state}\;,\nonumber\\
&&\textrm{if} \quad \lvert\vec\rho\rvert<1 \quad \textrm{mixed state}\;.
\end{eqnarray}
Explicitly, the density matrix for a pure state with spin $\vec\sigma$ along $\vec\alpha$
denotes
\begin{equation}\label{rho-up-alpha}
\rho=\lvert\Uparrow \vec\alpha\rangle\langle\Uparrow \vec\alpha\rvert=
    \left(\begin{array}{cc}
            \cos^2\frac{\alpha}{2} & \frac{1}{2}\sin\alpha\; e^{-i\phi} \\
            \frac{1}{2}\sin\alpha\; e^{i\phi} & \sin^2\frac{\alpha}{2} \\
    \end{array}\right)
\end{equation}\\
and for a mixed state with a $50:50$ mixture of spin up and down we have
\begin{equation}\label{rho-mixture}
\rho_{\rm mixed}=\frac{1}{2}\bigl(\lvert\Uparrow \rangle\langle\Uparrow \rvert +
\lvert\Downarrow \rangle\langle\Downarrow \rvert\bigr) = \frac{1}{2}\,\mathbbm{1}\;.
\end{equation}\\

\textbf{Proposition:} for a density matrix of mixed states
\begin{itemize}
\item [$\bullet$] \textit{There are different mixtures of states leading to the same
$\rho_{\rm mixed}$} !\\
\end{itemize}

\textbf{Example:}
\begin{equation*}
    \rho_{\rm mixed}\;=\;\frac{1}{2}\;\uparrow\downarrow \;\; = \;\frac{1}{3}\qquad\quad =
    \frac{1}{2}\,\mathbbm{1}
\end{equation*}
\setlength{\unitlength}{1cm}
\begin{picture}(10,0.5)(-5.85,0)
    \put(1,1){\vector(0,1){0.5}}
    \put(1,1){\vector(-1,-1){0.36}}
    \put(1,1){\vector(1,-1){0.36}}
\end{picture}

Here, the up--down arrows denote the mixed state (\ref{rho-mixture}), where the weight is
$\frac{1}{2}\,$, and the three star--like arrows represent a mixture of three states
(\ref{rho-up-alpha}) with the angles $\alpha = 0^{\textrm{o}}, \pm 120^{\textrm{o}}$
($\phi = 0^{\textrm{o}}$) and the weight $\frac{1}{3}\,$.\\

\textbf{Physics:}\\
\rule{0.2cm}{0cm}$\bullet$ $\;$ \textit{The physics depends only on the density matrix
$\rho$} !\\

\noindent $\Longrightarrow$ Several types of mixtures of the same $\rho_{\rm mixed}$ are
not distinguishable.\\
\rule{0.55cm}{0cm} They are different expressions of incomplete information about system.\\

\noindent $\Longrightarrow$ The entropy of a quantum system measures the degree of
uncertainty,\\
\rule{0.6cm}{0cm} i.e., the lack of knowledge, of the quantum state of a system.

\subsection{Model}
\label{Model}

We discuss the model of decoherence in a 2-dimensional Hilbert space
$\mathcal{H}=\mathbf{C}^2$ and consider the usual non-Hermitian ``effective mass''
Hamiltonian $H$ which describes the decay properties and the strangeness oscillations of
the kaons. The mass eigenstates, the short--lived $|K_S\rangle$ and long--lived
$|K_L\rangle$ states, are determined by
\begin{eqnarray}\label{Hamiltonian}
&&H\;|K_{S,L}\rangle\;=\;\lambda_{S,L}\;|K_{S,L}\rangle\qquad\textrm{with}\quad
\lambda_{S,L}\;=\;m_{S,L}-\frac{i}{2}\Gamma_{S,L} \;,
\end{eqnarray}
with $m_{S,L}$ and $\Gamma_{S,L}$ being the corresponding masses and decay widths. For
our purpose $CP$ invariance\footnote{Note that corrections due to $CP$ violations are of
order $10^{-3}$, however, we compare this model of decoherence with the data of the
CPLEAR experiment \cite{CPLEAR-EPR} which are not sensitive to $CP$ violating effects.}
is assumed, i.e. the $CP$ eigenstates $|K_1^0\rangle , |K_2^0\rangle$ are equal to the
mass eigenstates
\begin{eqnarray}\label{CPinvariance}
|K_1^0\rangle\equiv|K_S\rangle, \quad |K_2^0\rangle\equiv|K_L\rangle , \qquad\textrm{and}
\quad \langle K_S|K_L\rangle=0 \;.
\end{eqnarray}

As a starting point for our model of decoherence we consider the Liouville -- von Neumann
equation with the Hamiltonian (\ref{Hamiltonian}) and allow for decoherence by adding a
so--called \textit{dissipator} $D[\rho]$, so that the time evolution of the density
matrix $\rho$ is governed by the following \textit{master equation}
\begin{eqnarray}\label{Lindbladequation}
\frac{d\rho}{dt}\; &=&\; -\,i H\rho \,+\,i\rho H^\dagger\,-\,D[\rho] \;.
\end{eqnarray}
For the dissipative term $D[\rho]$ we write the simple ansatz (see
Refs.\cite{BG3,BertlmannDurstbergerHiesmayr2002})
\begin{eqnarray}\label{Dissipationterm}
D[\rho] \;=\; \lambda \, \big(P_S \rho P_L + P_L \rho P_S\big) \;=\; \frac{\lambda}{2}
\sum_{j=S,L} \big[P_j,[P_j,\rho]\big] \;,
\end{eqnarray}
where $P_j\;=\;|K_j\rangle\langle K_j|$ ($j=S,L$) denote the projectors to the
eigenstates of the Hamiltonian and $\lambda$ is called \textit{decoherence parameter};
$\lambda \ge 0$.\\

\textbf{Remark:} Note that we focus here on the undecayed kaon system which is described
by the non--Hermitian ``effective mass'' Hamiltonian (\ref{hamiltonian}),
(\ref{Hamiltonian}). In this case the master equation (\ref{Lindbladequation}) is not
trace conserving. But it, clearly, becomes trace conserving again when we include the
decay product states as well. So the total system is described by an enlarged Hilbert
space being the direct sum of the kaon-- and decay product space. Since our interest is
the decoherence study of the undecayed $K$--meson system we confine ourselves to master
equation (\ref{Lindbladequation}) neglecting the part of the decay products.\\

\textbf{Features:} With ansatz (\ref{Dissipationterm}) our model has the following nice
features.
\begin{enumerate}
\item[i)] It describes a completely positive map; when identifying
$A_j = \sqrt{\lambda} \, P_j \; , \; j=S,L$, it is a special case of Lindblad's general
structure \cite{Lindblad}
\begin{eqnarray}
D[\rho] \;=\; \frac{1}{2} \sum_j (A^\dagger_j A_j\; \rho+\rho A^\dagger_j A_j-2 A_j \rho
A_j^\dagger) \;.
\end{eqnarray}
Equivalently, it is a special form of the Gorini--Kossakowski--Sudarshan expression
\cite{GoriniKossakowskiSudarshan} (see Ref.\cite{BG4}).
\item[ii)] It conserves energy in case of a Hermitian Hamiltonian
since $[P_j,H]=0$ (see Ref.\cite{Adler1}).
\item[iii)] The von Neumann entropy $S(\rho)\,=\,-\;\mathrm{Tr}(\rho\ln\rho)$ is not
decreasing as a function of time since $P_j^\dagger = P_j$, thus $A_j^\dagger = A_j$ in
our case (see Ref.\cite{BenattiNarnhofer}).
\end{enumerate}

With choice (\ref{Dissipationterm}) the time evolution (\ref{Lindbladequation}) decouples
for the components of the matrix $\rho$, which are defined by
\begin{equation}
\rho (t) \;= \sum_{i,j=S,L} \rho_{ij} (t) \, | K_i \rangle \langle K_j | \;,
\end{equation}
and we obtain
\begin{eqnarray}\label{lambdatimeevolution}
\rho_{SS}(t)&=&\rho_{SS}(0)\cdot e^{-\Gamma_S t} \;,\nonumber\\
\rho_{LL}(t)&=&\rho_{LL}(0)\cdot e^{-\Gamma_L t} \;,\nonumber\\
\rho_{LS}(t)&=&\rho_{LS}(0)\cdot e^{-i \Delta m t - \Gamma t - \lambda t} \;,
\end{eqnarray}
where $\Delta m=m_L-m_S\,$ and $\Gamma=\frac{1}{2}(\Gamma_S+\Gamma_L)\,$.

\subsection{Entangled kaons}
\label{Entangled kaons}

Let us study now entangled neutral kaons. We use the abbreviations
\begin{eqnarray}\label{2-states}
|e_1\rangle\;=\;|K_S\rangle_l\otimes|K_L\rangle_r\qquad\textrm{and}\qquad
|e_2\rangle\;=\;|K_L\rangle_l\otimes|K_S\rangle_r \;,
\end{eqnarray}
and regard ---as usual--- the total Hamiltonian as a tensor product of the 1--particle
Hilbert spaces: $H=H_l\otimes\mathbf{1}_r + \mathbf{1}_l\otimes H_r$, where $l$ denotes
the left--moving and $r$ the right--moving particle. The initial Bell singlet state
\begin{eqnarray}\label{singletstate}
|\psi^{-}\rangle&=&\frac{1}{\sqrt{2}}\biggl\lbrace |e_1\rangle-|e_2\rangle\biggr\rbrace
\end{eqnarray}
is equivalently described by the density matrix
\begin{eqnarray}\label{rhozero}
\rho(0)\;=\;|\psi^{-}\rangle\langle \psi^{-}|\;=\;\frac{1}{2}\biggl\lbrace
|e_1\rangle\langle e_1| +|e_2\rangle\langle e_2|-|e_1\rangle\langle
e_2|-|e_2\rangle\langle e_1|\biggr\rbrace\;.
\end{eqnarray}
Then the time evolution given by (\ref{Lindbladequation}) with our ansatz
(\ref{Dissipationterm}), where now the operators $P_j\;=\;|e_j\rangle\langle e_j|$
($j=1,2$) project to the eigenstates of the 2--particle Hamiltonian $H$, also decouples
\begin{eqnarray}
\rho_{11}(t)&=&\rho_{11}(0)\; e^{-2 \Gamma t} \;,\nonumber\\
\rho_{22}(t)&=&\rho_{22}(0)\; e^{-2 \Gamma t} \;,\nonumber\\
\rho_{12}(t)&=&\rho_{12}(0)\, e^{-2 \Gamma t-\lambda t} \;.
\end{eqnarray}
Consequently, we obtain for the time-dependent density matrix
\begin{eqnarray}\label{matrixevolutionsolution}
\rho(t) \;=\; \frac{1}{2} e^{-2 \Gamma t} \biggl\lbrace  |e_1\rangle\langle e_1| +
|e_2\rangle\langle e_2| - e^{-\lambda t} \big( \,|e_1\rangle\langle e_2| +
|e_2\rangle\langle e_1| \, \big) \biggr\rbrace \;.
\end{eqnarray}
The decoherence arises through the factor $e^{-\lambda t}$ which only effects the
off--diagonal elements. It means that for $t>0$ and $\lambda \not = 0$ the density matrix
$\rho(t)$ does not correspond to a pure state anymore.

Finally, in order to arrive at a proper density matrix for the kaon system, conditioned
on having not decayed up to time $t\,$, we have to divide $\rho(t)$
(\ref{matrixevolutionsolution}) by the trace $\mathrm{Tr}\rho(t)\,$, see Sect.
\ref{Von-Neumann-entropy}.

\subsection{Measurement}
\label{Measurement}

In our model the parameter $\lambda$ quantifies the strength of possible decoherence of
the whole entangled state. We want to determine its permissible range of values by
experiment.

Concerning the measurement, we have the following point of view. The 2--particle density
matrix follows the time evolution given by Eq.(\ref{Lindbladequation}) with the Lindblad
generators $A_j=\sqrt{\lambda}\; |e_j\rangle\langle e_j|$ and undergoes thereby some
decoherence. We measure the strangeness content $S$ of the right--moving particle at time
$t_r$ and of the left--moving particle at time $t_l$. For sake of definiteness we choose
$t_r\leq t_l$. For times $t_r\leq t\leq t_l$ we have a 1--particle state which evolves
exactly according to QM, i.e. no further decoherence is picked up.

In theory we describe the measurement of the strangeness content, i.e. the right--moving
particle being a $K^0$ or a $\bar K^0$ at time $t_r$, by the following projection onto
$\rho$
\begin{eqnarray}
\mathrm{Tr}_r \,\big\{ \mathbf{1}_l \otimes |S^{'}\rangle\langle S^{'}|_r \;\; \rho(t_r)
\big\} \;&\equiv&\; \rho_l(t=t_r;t_r) \;,
\end{eqnarray}
where strangeness $S^{'} = +,-$ and $|+\rangle = |K^0\rangle$, $|-\rangle = |\bar
K^0\rangle$. Consequently, $\rho_l(t;t_r)$ for times $t \ge t_r$ is the 1--particle
density matrix for the left--moving particle and evolves as a 1--particle state according
to pure QM. At $t=t_l$ the strangeness content ($S=+,-$) of the second particle is
measured and we finally calculate the probability
\begin{eqnarray}
P_\lambda(S,t_l;S^{'},t_r)\;&=&\;\mathrm{Tr}_l\,\big\{|S\rangle\langle
S|_l\;\;\rho_l(t_l;t_r)\big\}\;.
\end{eqnarray}
Explicitly, we find the following results for the like-- and unlike--strangeness
probabilities
\begin{eqnarray}\label{lambdaprobabilities}
\lefteqn{P_\lambda(K^0,t_l;K^0,t_r)\;=\;P_\lambda(\bar K^0,t_l;\bar K^0,t_r)}\nonumber\\
&=&\frac{1}{8} \biggl\lbrace e^{-\Gamma_S t_l-\Gamma_L t_r}+e^{-\Gamma_L t_l-\Gamma_S
t_r}-e^{-\lambda t_r}\;2
\cos(\Delta m \Delta t)\cdot e^{-\Gamma (t_l+t_r)}\biggr\rbrace\;,\nonumber\\
\lefteqn{P_\lambda(K^0,t_l;\bar K^0,t_r)\;=\;P_\lambda(\bar K^0,t_l;K^0,t_r)}\nonumber\\
&=&\frac{1}{8} \biggl\lbrace e^{-\Gamma_S t_l-\Gamma_L t_r}+e^{-\Gamma_L t_l-\Gamma_S
t_r}+e^{-\lambda t_r}\;2 \cos(\Delta m \Delta t)\cdot e^{-\Gamma (t_l+t_r)}\biggr\rbrace
\end{eqnarray}
with $\Delta t = t_l-t_r$.

Note that at equal times $t_l=t_r=t$ the like--strangeness probabilities
\begin{equation}
P_\lambda(K^0,t;K^0,t)\;=\;P_\lambda(\bar K^0,t;\bar K^0,t)\;=\; \frac{1}{4} \;
e^{-2\Gamma t} \; (1 - e^{-\lambda t})
\end{equation}
do not vanish, in contrast to the pure quantum mechanical EPR-correlations.\\

The interesting quantity is the \textit{asymmetry of probabilities}; it is directly
sensitive to the interference term and can be measured experimentally. For pure QM we
have
\begin{eqnarray}\label{qmasymmetry}
\lefteqn{A^{QM}(\Delta t) =} \nonumber\\
=&& \frac{P(K^0,t_l;\bar K^0,t_r)+P(\bar K^0,t_l;K^0,t_r)-P(K^0,t_l;K^0,t_r) -P(\bar
K^0,t_l;\bar K^0,t_r)} {P(K^0,t_l;\bar K^0,t_r)+P(\bar K^0,t_l; K^0, t_r)
+P(K^0,t_l;K^0,t_r)+P(\bar K^0,t_l;\bar K^0,t_r)}\nonumber\\
=&& \;\frac{\cos(\Delta m \Delta t)}{\cosh(\frac{1}{2}\Delta \Gamma \Delta t)} \;,
\end{eqnarray}
with $\Delta\Gamma = \Gamma_L-\Gamma_S$, and for our decoherence model we find, by
inserting the probabilities (\ref{lambdaprobabilities}),
\begin{eqnarray}\label{lambdaasymmetry}
A^\lambda(t_l,t_r)\;&=&\;\frac{\cos(\Delta m \Delta t)}{\cosh(\frac{1}{2}\Delta
\Gamma\Delta t)} \;\; e^{-\lambda \min{\{t_l,t_r\}}}\nonumber\\
\;&=&\; A^{QM}(\Delta t) \;\; e^{-\lambda \min{\{t_l,t_r\}}} \;.
\end{eqnarray}
Thus the decoherence effect, simply given by the factor $e^{-\lambda \min{\{t_l,t_r\}}}$,
depends only ---according to our philosophy--- on the time of the first measured kaon, in
our case: $\min{\{t_l,t_r\}} = t_r$.

\subsection{Experiment}
\label{Experiment}

Now we compare our model with the results of the CPLEAR experiment \cite{CPLEAR-EPR} at
CERN where $K^0 \bar K^0$ pairs are produced in the $p \bar p$ collider: $p\bar
p\longrightarrow K^0 \bar K^0$. These pairs are predominantly in an antisymmetric state
with quantum numbers $J^{PC}=1^{- -}$ and the strangeness of the kaons is detected via
strong interactions in surrounding absorbers (made of copper and carbon).\\

\textbf{Examples:}
\begin{eqnarray*}
    S=+1:\qquad K^0(d\bar s)+N&\longrightarrow &K^+(u\bar s)+ X\;,\\
    S=-1:\qquad \bar K^0(\bar d s)+N&\longrightarrow &K^-(\bar u s)+ X\;,\\
    S=-1:\qquad \bar K^0(\bar d s)+N&\longrightarrow &\Lambda(uds)+ X \quad \mbox{and}\quad
    \Lambda\longrightarrow p\pi^-\;;
\end{eqnarray*}
like--strangeness events: $(K^-,\Lambda)\;$,\\
unlike--strangeness events: $(K^+,K^-)$, $(K^+,\Lambda)\;$.\\

\begin{figure}
\center{\includegraphics[width=200pt, height=150pt]{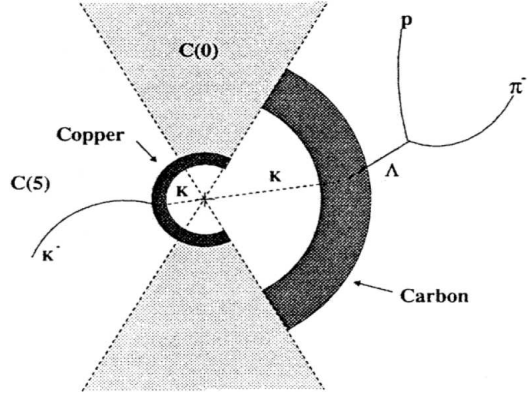}}

\caption{Example of CPLEAR event: like-strangeness $(K^-,\Lambda)$}\label{setup}
\end{figure}

The experimental set--up has two configurations (see Fig.\ref{setup}). In configuration
$C(0)$ both kaons propagate $2$ cm, they have nearly equal proper times ($t_r\approx
t_l$) when they are measured by the absorbers. This fulfills the condition for an
EPR--type experiment. In configuration $C(5)$ one kaon propagates $2$ cm and the other
kaon $7$ cm, thus the flight--path difference is $5$ cm on average, corresponding to a
proper time difference $|t_r-t_l|\approx 1.2 \tau_S\,$.

\begin{figure}
\center{\includegraphics[width=200pt, height=150pt]{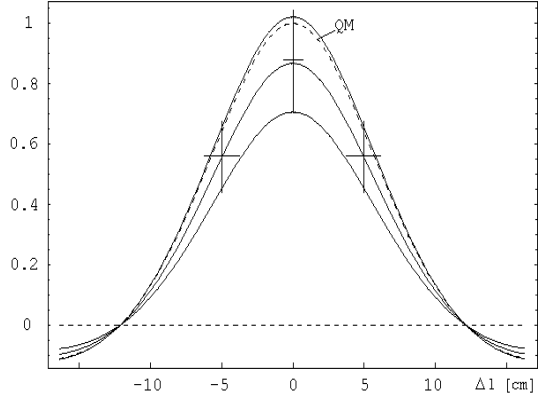}}

\caption{The asymmetry (\ref{lambdaasymmetry}) as a function of the flight--path
difference of the kaons. The dashed curve corresponds to QM, the solid curves represent
the best fit (\ref{lambda-values}) to the CPLEAR data \cite{CPLEAR-EPR}, given by the
crosses. The horizontal line indicates the Furry--Schr\"odinger's hypothesis
\cite{Schrodinger,Furry}, as explained in Sect.
\ref{connection-to-pheno-model}.}\label{deco-fit}
\end{figure}

Fitting the decoherence parameter $\lambda$ by comparing the asymmetry
(\ref{lambdaasymmetry}) with the experimental data \cite{CPLEAR-EPR} (see
Fig.\ref{deco-fit}) we find, when averaging over both configurations, the following
bounds on $\lambda$
\begin{eqnarray}\label{lambda-values}
\bar\lambda \;=\; (1.84^{+2.50}_{-2.17})\cdot 10^{-12}\;\textrm{MeV}
\quad\textrm{and}\quad \bar\Lambda \;=\;
\frac{\bar\lambda}{\Gamma_S}\;=\;0.25^{+0.34}_{-0.32}\;.
\end{eqnarray}
The results (\ref{lambda-values}) are certainly compatible with QM ($\lambda = 0$),
nevertheless, the experimental data allow an upper bound $\bar\lambda_{\textrm{up}} =
4.34 \cdot 10^{-12}\;\textrm{MeV}$ for possible decoherence in the entangled $K^0 \bar
K^0$ system.\\

\textbf{$B^0 \bar B^0$ system:} An analogous investigation of ``decoherence of entangled
beauty'' has been carried out in Ref.\cite{BG3}. However, there time integrated events
from the semileptonic decays of the entangled $B^0 \bar B^0$ pairs are analyzed, which
were produced at the colliders at DESY and Cornell. The analysis provides the bounds
$\lambda_B = (-47\pm76)\cdot 10^{-12}\;\textrm{MeV}$, which are an order of magnitude
less restrictive than the bounds (\ref{lambda-values}) of the $K^0 \bar K^0$ system.

The possibility to measure the asymmetry at different times is offered now also in the
$B$--meson system. Entangled $B^0\bar B^0$ pairs are created with high density at the
asymmetric $B$--factories and identified by the detectors BELLE at KEK-B (see e.g. Refs.
\cite{Belle,Leder}) and BABAR at PEP-II (see e.g. Refs. \cite{Aubert,Babar}) with a high
resolution at different distances or times.

\section{Connection to phenomenological model}
\label{connection-to-pheno-model}

There exists a simple procedure \cite{BG1,Dass,BG2} for introducing decoherence in a
rather phenomenological way, more in the spirit of Furry \cite{Furry} and Schr\"odinger
\cite{Schrodinger} to describe the process of spontaneous factorization of the
wavefunction. QM is modified in the sense that we multiply the interference term of the
transition amplitude by the factor $(1-\zeta)$. The quantity $\zeta$ is called
\textit{effective decoherence parameter}. Starting again from the Bell singlet state
$|\psi^- \rangle$, which is given by the mass eigenstate representation
(\ref{singletstate}), we have for the like--strangeness probability
\begin{eqnarray}\label{plikezeta}
\lefteqn{P(K^0,t_l;K^0,t_r)\;=\;||\langle K^0|_l \otimes \langle K^0|_r \;
|\psi^-(t_l,t_r)\rangle||^2}\nonumber\\
&\longrightarrow& \quad P_{\zeta}(K^0, t_l; K^0, t_r) \;=\; \frac{1}{2}\;\biggl\lbrace
e^{-\Gamma_S t_l-\Gamma_L t_r} |\langle K^0|K_S\rangle_l|^2\;
|\langle K^0|K_L\rangle_r|^2\nonumber\\
& &+\; e^{-\Gamma_L t_l-\Gamma_S t_r}|\langle K^0|K_L\rangle_l|^2\; |\langle
K^0|K_S\rangle_r|^2  \;-\; 2\; \underbrace{(1-\zeta)}\; e^{-\Gamma(t_l+t_r)}\nonumber\\
& &\quad\hspace{6.2cm}\textrm{modification} \nonumber\\
& &\times\;\mathrm{Re}\big\{\langle K^0|K_S\rangle_l^* \langle K^0|K_L\rangle_r^* \langle
K^0|K_L\rangle_l \langle K^0|K_S\rangle_r\;
e^{-i \Delta m \Delta t}\big\} \biggr\rbrace \nonumber\\
&=&\frac{1}{8}\,\biggl\lbrace e^{-\Gamma_S t_l-\Gamma_L t_r} + e^{-\Gamma_L t_l-\Gamma_S
t_r} \;-\; 2\, \underbrace{(1-\zeta)}\, e^{-\Gamma(t_l+t_r)}\, \cos(\Delta m \Delta
t)\biggr\rbrace\,,
\nonumber\\
& &\hphantom{\frac{1}{8}\biggl\lbrace e^{-\Gamma_S t_l-\Gamma_L t_r} + e^{-\Gamma_L
t_l-\Gamma_S t_r}} \;\;\quad\textrm{modification}
\end{eqnarray}
and the unlike-strangeness probability just changes the sign of the interference term.\\

\textbf{Features:} The value $\zeta=0$ corresponds to pure QM and $\zeta=1$ to total
decoherence or spontaneous factorization of the wave function, which is commonly known as
Furry--Schr\"odinger hypothesis \cite{Schrodinger,Furry}. The decoherence parameter
$\zeta$, introduced in this way ``by hand'', is quite effective \cite{BG1,Dass,BG2,BGH};
it interpolates continuously between the two limits: QM $\longleftrightarrow$ spontaneous
factorization. It represents a measure for the amount of decoherence which results in a
loss of entanglement of the total quantum state (we come back to this point
in Section \ref{Entanglement-Measure}).\\

There exists a remarkable one--to--one correspondence between the model of decoherence
(\ref{Lindbladequation}), (\ref{Dissipationterm}) and the phenomenological model, thus a
relation between $\lambda \longleftrightarrow \zeta$
(Refs.\cite{BG3,BertlmannDurstbergerHiesmayr2002}). We can see it quickly in the
following way.\\

Calculating the asymmetry of strangeness events, as defined in Eq.(\ref{qmasymmetry}),
with the probabilities (\ref{plikezeta}) we obtain
\begin{eqnarray}\label{zetaasymmetry}
A^{\zeta}(t_l,t_r) \;=\; A^{QM}(\Delta t) \; \big(\,1 \,-\, \zeta(t_l,t_r)\,\big)\;.
\end{eqnarray}
When we compare now the two approaches, i.e. Eq.(\ref{lambdaasymmetry}) with
Eq.(\ref{zetaasymmetry}), we find the formula
\begin{equation}\label{zetamin}
\zeta(t_l,t_r) \;=\;1 \,-\; e^{-\lambda \min{\{t_l,t_r\}}} \;.
\end{equation}
Of course, when fitting the experimental data with the two models, the $\lambda$ values
(\ref{lambda-values}) are in agreement with the corresponding $\zeta$ values (averaged
over both experimental setups) \cite{BGH,trixi},
\begin{equation}\label{zeta-values}
\bar\zeta \;=\; 0.13^{+0.16}_{-0.15}\;.
\end{equation}

The phenomenological model demonstrates that the $K^0 \bar K^0$ system is close to QM,
i.e. $\zeta=0$, \textit{and} (at the same time) far away from total decoherence, i.e.
$\zeta=1$. It confirms nicely in a quantitative way the existence of entangled massive
particles over macroscopic distances ($9$) cm.\\

We consider the decoherence parameter $\lambda$ to be the fundamental constant, whereas
the value of the effective decoherence parameter $\zeta$ depends on the time when a
measurement is performed. In the time evolution of the state $|\psi^- \rangle$,
Eq.(\ref{singletstate}), represented by the density matrix
(\ref{matrixevolutionsolution}), we have the relation
\begin{equation}\label{zeta}
\zeta(t) \;=\; 1 \,-\; e^{-\lambda t} \;,
\end{equation}
which after measurement of the left-- and right--moving particles at $t_l$ and $t_r$
turns into formula (\ref{zetamin}), if decoherence occurs as described in Sect.
\ref{Measurement}.\\

\vspace{0.3cm}

Our model of decoherence has a very specific time evolution (\ref{zetamin}). Measuring
the strangeness content of the entangled kaons at definite times we have the possibility
to distinguish experimentally, on the basis of time dependent event rates, between the
prediction of our model (\ref{zetamin}) and the results of other models (which would
differ from Eq.(\ref{zetamin})). Indeed, it is of high interest to measure in future
experiments the asymmetry of the strangeness events at various times, in order to confirm
the very specific time dependence of the decoherence effect. In fact, such a possibility
is now offered in the $B$--meson system; as we already mentioned entangled $B^0\bar B^0$
pairs are created with high density at the asymmetric $B$--factories at KEK-B and at
PEP-II.


\section{Entanglement loss --- decoherence}
\label{Entanglement-Measure}

In the master equation (\ref{Lindbladequation}) the \textit{dissipator} $D[\rho]$
describes two phenomena occurring in an open quantum system, namely decoherence and
dissipation (see, e.g., Ref. \cite{BreuerPetruccione}). When the system $S$ interacts
with the environment $E$ the initially product state evolves into entangled states of
$S+E$ in the course of time \cite{Joos,KublerZeh}. It leads to mixed states in $S$
---which means decoherence--- and to an energy exchange between $S$ and $E$ ---which is
called dissipation.

The decoherence destroys the occurrence of long--range quantum correlations by
suppressing the off-diagonal elements of the density matrix in a given basis and leads to
an information transfer from the system $S$ to the\\
environment $E$: $\hspace{2.0cm}$ \fbox{\fbox{S}$\rightleftarrows$ \hspace{0.4cm} E}\\

In general, both effects are present, however, decoherence acts on a much shorter time
scale \cite{Joos,JoosZeh,Zurek,Alicki} than dissipation and is the more important effect
in quantum information processes.

Our model describes decoherence and not dissipation. The increase of decoherence of the
initially totally entangled $K^0 \bar  K^0$ system as time evolves means on the other
hand a decrease of entanglement of the system. This loss of entanglement can be measured
explicitly \cite{BertlmannDurstbergerHiesmayr2002,PHDtrixi}.

In the field of quantum information the entanglement of a state is quantified by
introducing certain \textit{entanglement measures}. In this connection the entropy plays
a fundamental role.\\

\textbf{Entropy of a quantum system:}
\begin{itemize}
    \item [$\bullet$] \textit{The entropy measures the degree of uncertainty ---the lack
    of knowledge--- of a quantum state} !
\end{itemize}

In general, a quantum state can be in a pure or mixed state (see, our discussion in Sect.
\ref{Density matrix}). Whereas the pure state supplies us with maximal information about
the system, a mixed state does not.\\

\textbf{Proposition:}$\;$ Thirring  \cite{Thirring}
\begin{itemize}
    \item [$\bullet$] \textit{Mixed states provide only partial information about the
    system; the entropy measures how much of the maximal information is missing} !
\end{itemize}

For mixed states the entanglement measures cannot be defined uniquely. Common measures of
entanglement are von Neumann's entropy function $S$, the entanglement of formation $E$
and the concurrence $C$.

\subsection{Von Neumann entropy}
\label{Von-Neumann-entropy}

We are only interested in the effect of decoherence, thus we properly normalize the state
(\ref{matrixevolutionsolution}) in order to compensate for the decay property
\begin{eqnarray}\label{normdensitymatrix}
\rho_N(t) &=& \frac{\rho(t)}{\mathrm{Tr}\rho(t)} \;.
\end{eqnarray}
\textit{Von Neumann's entropy} for the quantum state (\ref{normdensitymatrix}) is defined
by
\begin{eqnarray}\label{vonNeumannentropy}
S\big(\rho_N(t)\big) \;&=&\; -\; \mathrm{Tr}\{\rho_N(t)\log_2 \rho_N(t)\}\nonumber\\
\;&=&\; - \frac{1-e^{-\lambda t}}{2}\log_2 \frac{1-e^{-\lambda t}}{2} \;-\;
\frac{1+e^{-\lambda t}}{2}\log_2 \frac{1+e^{-\lambda t}}{2} \;,\hspace{0.5cm}
\end{eqnarray}
where we have chosen the logarithm to base $2$, the dimension of the Hilbert space
(the qubit space), such that $S$ varies between $0\leq S \leq 1\,$.\\

\textbf{Features:}
\begin{enumerate}
\item[i)] $S\big(\rho_N(t)\big) = 0$ for $t=0\,$; the entropy is zero at time $t=0\,$, there
is no uncertainty in the system, the quantum state is pure and maximally entangled.
\item[ii)] $S\big(\rho_N(t)\big) = 1$ for $t \to \infty$; the entropy increases for
increasing $t$ and approaches the value $1$ at infinity. Hence the state becomes
more and more mixed.\\
\end{enumerate}

\textbf{Reduced density matrices:} $\;$ Let us consider quite generally a composite
quantum system $A$ (Alice) and $B$ (Bob). Then the reduced density matrix of subsystem
$A$ is given by tracing the density matrix of the joint state over all states of $B$.

In our case the subsystems are the propagating kaons on the left $l$ and right $r$ hand
side thus we have as \textit{reduced density matrices}
\begin{equation}\label{reduceddensity}
\rho^{\,l}_N(t) = \mathrm{Tr}_r\{\rho_N(t)\} \quad \textrm{and} \quad \rho^{\,r}_N(t) =
\mathrm{Tr}_l\{\rho_N(t)\} \;.
\end{equation}

The \textit{uncertainty in the subsystem} $l$ before the subsystem $r$ is measured is
given by the von Neumann entropy $S(\rho^{\,l}_N(t))$ of the corresponding reduced
density matrix $\rho^{\,l}_N(t)$ (and alternatively we can replace $l \to r$). In our
case we find
\begin{equation}\label{entropysubsystems}
S\big(\rho^{\,l}_N(t)\big) = S\big(\rho^{\,r}_N(t)\big) = 1  \;\qquad \forall \; t \geq 0
\;.
\end{equation}
\textit{The reduced entropies are independent of $\lambda$}! That means the correlation
stored in the composite system is, with increasing time, lost into the environment
---what intuitively we had expected--- and \textit{not} into the subsystems, i.e. the
individual kaons.

For pure quantum states von Neumann's entropy function (\ref{vonNeumannentropy}) is a
good measure for entanglement and, generally, $A$ (Alice) and $B$ (Bob) are most
entangled when their reduced density matrices are maximally mixed.

For mixed states, however, von Neumann's entropy and the reduced entropies are no longer
a good measure for entanglement so that we have to proceed in an other way to quantify
entanglement (see Sect. \ref{Entanglement-formation-concurrence}).

\subsection{Separability}
\label{separability}

In the following we want to show that the initially entangled Bell singlet state
---although subjected to decoherence and thus to entanglement loss in the course of
time--- remains entangled. It is convenient to work with the ``quasi--spin'' description
for the $K^0 \bar  K^0$ system (see Sect. \ref{SpinAnalogy}). The projection operators of
the mass eigenstates correspond to the spin projection operators ``up'' and ``down''
\begin{eqnarray}
P_S &=& |K_S\rangle\langle K_S| \;=\; \sigma_{\uparrow} \;=\; \frac{1}{2} \, (\mathbf{1}
+ \sigma_z) \;=\; \left( \ba{cc}
1 & 0\\
0 & 0\\
\ea \right) \;,\nonumber\\
P_L &=& |K_L\rangle\langle K_L| \;=\; \sigma_{\downarrow} \;=\; \frac{1}{2} \,
(\mathbf{1} - \sigma_z) \;=\; \left( \ba{cc}
0 & 0\\
0 & 1\\
\ea \right) \;,
\end{eqnarray}
and the transition operators are the ``spin--ladder'' operators
\begin{eqnarray}
P_{SL} &=& |K_S\rangle\langle K_L| \;=\; \sigma_{+} \;=\; \frac{1}{2} \, (\sigma_x + i \,
\sigma_y) \;=\; \left( \ba{cc}
0 & 1\\
0 & 0\\
\ea \right) \;,\nonumber\\
P_{LS} &=& |K_L\rangle\langle K_S| \;=\; \sigma_{-} \;=\; \frac{1}{2} \, (\sigma_x - i \,
\sigma_y) \;=\; \left( \ba{cc}
0 & 0\\
1 & 0\\
\ea \right) \;.
\end{eqnarray}
Then density matrix (\ref{normdensitymatrix}), (\ref{matrixevolutionsolution}) is
expressed by the Pauli spin matrices in the following way
\begin{equation}\label{densitymatrixspinxyz}
\rho_N(t) \,=\, \frac{1}{4}\big\{\mathbf{1} - \sigma_z\otimes\sigma_z \,-\, e^{-\lambda
t}\, [\sigma_x\otimes\sigma_x + \sigma_y\otimes\sigma_y]\big\} \;,
\end{equation}
which at $t=0$ coincides with the well-known expression for the pure spin singlet state
$\rho_N(t=0) \,=\, \frac{1}{4} \, (\mathbf{1} - \vec\sigma\otimes\vec\sigma)$; see, e.g.,
Ref. \cite{BNT}.

Operator (\ref{densitymatrixspinxyz}) can be nicely written as $4 \times 4$ matrix
\begin{eqnarray}\label{densitymatrix4x4}
\rho_N(t) \;=\; \frac{1}{2} \left( \ba{cccc}
0 & 0 & 0 & 0\\
0 & 1 & -e^{-\lambda t} & 0\\
0 & -e^{-\lambda t} & 1 & 0\\
0 & 0 & 0 & 0\\
\ea \right) \;.
\end{eqnarray}

For an other representation of the density matrix $\rho_N(t)$ we choose the so--called
``Bell basis''
\begin{eqnarray}
\rho^{\mp} \;=\; |\psi^{\mp}\rangle\langle \psi^{\mp}| \qquad\textrm{and}\qquad
\omega^{\mp} \;=\; |\phi^{\mp}\rangle\langle \phi^{\mp}| \;,
\end{eqnarray}
with $|\psi^{-}\rangle$ given by Eq.(\ref{singletstate}) and  $|\psi^{+}\rangle$ by
\begin{eqnarray}\label{symmetricstate}
|\psi^{+}\rangle&=&\frac{1}{\sqrt{2}}\biggl\lbrace |e_1\rangle+|e_2\rangle\biggr\rbrace
\;.
\end{eqnarray}
The states $|\phi^{\mp}\rangle = \frac{1}{\sqrt{2}} (|\Uparrow\Uparrow\rangle \mp
|\Downarrow\Downarrow\rangle)$ (in spin notation) do not contribute here.

\subsubsection{Entanglement --- separability}

Recall that in general the density matrix $\rho$ of a state is defined over the tensor
product of Hilbert spaces $\mathcal{H}=\mathcal{H}_A\otimes\mathcal{H}_B$, named Alice
and Bob.

A state $\rho$ is then called \textit{entangled} if it is \textit{not separable}, i.e.
$\;\rho\in S^c$ where $S^c$ is the complement of the set of separable states $S$; and
$S\cup S^c=\mathcal{H}$ .\\

\textbf{Separable states:} $\;$ The set of separable states is defined by
\begin{equation}
    S \;=\; \Bigl\{ \rho\, = \sum_{i} p_{i} \, \rho_A^i \otimes \rho_B^i \;\; \big\vert \;\;
    0 \leq p_{i} \leq 1, \;\sum_{i} p_{i} = 1\Bigr\} \;,
\end{equation}
where $\rho_A^i$ and $\rho_B^i$ are density matrices over the subspaces $\mathcal{H}_A$
and $\mathcal{H}_B$.\\

The important question is now to judge whether a quantum state is entangled or conversely
separable or not. Several \textit{separability criteria}  give an answer to that.
\begin{theorem}
\mbox{\rm{Positive partial transpose criterion,
Peres--Horodecki \cite{Peres,Horodecki2}}}\\
Defining the partial transposition $T_B$ by transposing only one of the subsystems, e.g.
$T_B (\sigma^i)_{kl} = (\sigma^i)_{lk} \,$ in subsystem $B$, then a state $\rho$ is
separable iff its partial transposition with respect to any subsystem is positive:
\begin{equation}\label{PPT-criterion}
(\mathbbm{1}_A\otimes T_B)\,\rho\,\geq\,0 \quad\textrm{and}\quad (T_A\otimes
\mathbbm{1}_B)\,\rho\,\geq\,0 \quad\; \Longleftrightarrow \quad\; \rho\;\;
\mbox{separable} \;.
\end{equation}
\end{theorem}
\vspace{0.02cm}
\begin{theorem}
\mbox{\rm{Reduction criterion, Horodecki \cite{Horodecki}}}\\
A state $\rho$ is separable for:
\begin{equation}\label{reduction-criterion}
\mathbbm{1}_A\otimes \rho_B - \,\rho\,\geq\,0 \quad\textrm{and}\quad \rho_A\otimes
\mathbbm{1}_B - \,\rho\,\geq\,0 \quad\; \Longleftrightarrow \quad\; \rho\;\;
\mbox{separable} \;,
\end{equation}
where $\rho_A$ is Alice's reduced density matrix and $\rho_B$ Bob's.\\
\end{theorem}

However, above Theorems (\ref{PPT-criterion}), (\ref{reduction-criterion}) are necessary
and sufficient separability conditions ---and so surprisingly simple--- only for
dimensions $2\otimes 2$ and $2\otimes 3$ \cite{HHHbook}. A more general
separability---entanglement criterion, valid in any dimensions, does exist; it is
formulated by a so--called \textit{generalized Bell inequality}, see Ref.\cite{BNT}.\\

\vspace{0.15cm}

Now let us return to the question of entanglement and separability of our kaon quantum
state described by density matrix $\rho_N(t)$ (\ref{normdensitymatrix}),
(\ref{matrixevolutionsolution}) as it evolves in time.\\

\newpage

\textbf{Proposition:}$\;$ Bertlmann--Durstberger--Hiesmayr
\cite{BertlmannDurstbergerHiesmayr2002}
\begin{itemize}
\item [$\bullet$] \textit{The state represented by the density matrix $\rho_N(t)$
(\ref{normdensitymatrix}), (\ref{matrixevolutionsolution}) becomes mixed for $0<t<\infty$
but remains entangled. Separability is achieved asymptotically $t \to \infty$ with the
weight $e^{-\lambda t}$. Explicitly, $\rho_N(t)$ is the following mixture of the Bell
states $\rho^{-}$ and $\rho^{+}\,$}:
\begin{equation}\label{densitymatrixBell}
\rho_N(t) \;=\; \frac{1}{2} \big(1 + e^{-\lambda t}\big) \, \rho^{-} \;+\; \frac{1}{2}
\big(1 - e^{-\lambda t}\big) \, \rho^{+} \;.\\
\end{equation}
\end{itemize}

\textbf{Proof:}
\begin{itemize}
\item[i)] The mixedness of the state, with $t\to\infty$ totally mixed --- separable,
can be seen from the \textit{mixed state criterion} (see Sect. \ref{Density matrix})
\begin{eqnarray}\label{densitymatrixsquared4x4}
\rho^{\,2}_N(t) \;=\; \frac{1}{4} \left( \ba{cccc}
0 & 0 & 0 & 0\\
0 & 1+e^{-2\lambda t} & -2e^{-\lambda t} & 0\\
0 & -2e^{-\lambda t} & 1+e^{-2\lambda t} & 0\\
0 & 0 & 0 & 0\\
\ea \right) \;\not=\; \rho_N(t) \quad \textrm{for} \quad t>0 \;.
\end{eqnarray}

\item[ii)] Entanglement or lack of separability is determined by Theorems
(\ref{PPT-criterion}) and (\ref{reduction-criterion}). The Peres--Horodecki partial
transposition criterion (\ref{PPT-criterion})
\begin{eqnarray}\label{rho-N-PPT}
(\mathbf{1}_l \otimes T_r) \, \rho_N(t) \;&=&\; \frac{1}{2} \left( \ba{clrc}
0 & 0 & 0 & -e^{-\lambda t}\\
0 & 1 & 0 & 0\\
0 & 0 & 1 & 0\\
-e^{-\lambda t} & 0 & 0 & 0\\
\ea \right) \not\geq 0 \;,
\end{eqnarray}
with eigenvalues $\big\{\frac{1}{2},\frac{1}{2},\frac{1}{2} e^{-\lambda t}, -\frac{1}{2}
e^{-\lambda t}\big\}\;$, is not positive. Alternatively, the Horodecki reduction
criterion (\ref{reduction-criterion}), a matrix with same eigenvalues (\ref{rho-N-PPT})
\begin{eqnarray}\label{rho-N-reduc}
\mathbf{1}_l\otimes\rho^{\,r}_N(t) - \rho_N(t) \;=\; \frac{1}{2} \left( \ba{cccc}
1 & 0 & 0 & 0\\
0 & 0 & e^{-\lambda t} & 0\\
0 & e^{-\lambda t} & 0 & 0\\
0 & 0 & 0 & 1\\
\ea \right) \not\geq 0 \;,
\end{eqnarray}
is not positive either. Therefore $\rho_N(t)$ remains entangled for $t<\infty$. $\Box$
\end{itemize}

\subsection{Entanglement of formation and concurrence}
\label{Entanglement-formation-concurrence}

For pure states the entropy of the reduced density matrices is sufficient, for mixed
states we need another measure, e.g., \textit{entanglement of formation}.

\subsubsection{Entanglement of formation}

Every density matrix $\rho$ can be decomposed into an ensemble of pure states
$\rho_i=|\psi_i\rangle\langle\psi_i|$ with the probability $p_i$, i.e. $\rho=\sum_i p_i
\rho_i$. The entanglement of formation for a pure state is given by the entropy of either
of the two subsystems. For a mixed state \textit{entanglement of formation}
\cite{Bennett-entangle} is defined as average entanglement of pure states of the
decomposition, minimized over all decompositions of $\rho$
\begin{eqnarray}\label{entanglementofformation}
E(\rho)&=&\min\sum_i \,p_i\, S(\rho_i^{\,l})\;.
\end{eqnarray}
It quantifies the resources needed to create a given entangled state. Bennett et al.
\cite{Bennett-entangle} found a remarkable simple formula for \textit{entanglement of
formation}
\begin{eqnarray}\label{entanglementbound}
E(\rho) &\ge& \mathcal{E}\big(f(\rho)\big) \;,
\end{eqnarray}
where the function $\mathcal{E}\big(f(\rho)\big)$ is defined by
\begin{eqnarray}\label{entanglementformula-f}
\mathcal{E}\big(f\big) \;=\; H\bigg(\frac{1}{2} + \sqrt{f(1-f)}\bigg) \quad \textrm{for}
\quad f \ge \frac{1}{2} \;,
\end{eqnarray}
and $\mathcal{E}\big(f\big) = 0$ for $f < \frac{1}{2}\,$. The function $H$ represents the
familiar binary entropy function $H(x) = -x\log_2x - (1-x)\log_2(1-x)$. The quantity
$f(\rho)$ is called the \textit{fully entangled fraction} of $\rho$
\begin{eqnarray}
f( \rho ) = \max \, \langle e|\rho|e \rangle \;,
\end{eqnarray}
which is the maximum over all completely entangled states $|e \rangle$.

For general mixed states $\rho$ the function $\mathcal{E}\big(f(\rho)\big)$ is only a
lower bound to the entropy $E(\rho)$. For pure states and mixtures of Bell states ---the
case of our model--- the bound is saturated, $E=\mathcal{E}$, and we have formula
(\ref{entanglementformula-f}) for calculating the entanglement of formation.

\subsubsection{Concurrence}

Wootters and Hill \cite{Wootters1,Wootters2,Wootters3} found that entanglement of
formation for a general mixed state $\rho$ of two qubits can be expressed by another
quantity, the \textit{concurrence} $C$
\begin{eqnarray}\label{entanglementformula-C}
E(\rho) \;&=&\; \mathcal{E}\big(C(\rho)\big) \;=\; H\bigg(\frac{1}{2} +
\frac{1}{2}\sqrt{1-C^2}\bigg) \quad \textrm{with} \quad 0 \le C \le 1 \;.
\end{eqnarray}
Explicitly, the function $\mathcal{E}(C)$ looks like
\begin{eqnarray}
\mathcal{E}(C)&=&-\frac{1+\sqrt{1-C^2}}{2}\log_2\frac{1+\sqrt{1-C^2}}{2}-
\frac{1-\sqrt{1-C^2}}{2}\log_2\frac{1-\sqrt{1-C^2}}{2}\nonumber\\
\end{eqnarray}
and is monotonically increasing from $0$ to $1$ as $C$ runs from $0$ to $1$. Thus $C$
itself is a kind of entanglement measure in its own right.

Defining the spin flipped state $\tilde\rho$ of $\rho$ by
\begin{eqnarray}
\tilde\rho \;&=&\; (\sigma_y\otimes\sigma_y)\, \rho^* (\sigma_y\otimes\sigma_y) \;,
\end{eqnarray}
where $\rho^*$ is the complex conjugate and is taken in the standard basis, i.e. the
basis $\{|\Uparrow\Uparrow\rangle, |\Downarrow\Downarrow\rangle,
|\Uparrow\Downarrow\rangle, |\Downarrow\Uparrow\rangle\}$, the \textit{concurrence} $C$
is given by the formula
\begin{eqnarray}
C(\rho) \;&=&\; \max\{0,\lambda_1-\lambda_2-\lambda_3-\lambda_4\} \;.
\end{eqnarray}
The $\lambda_i$'s are the square roots of the eigenvalues, in decreasing order, of the
matrix $\rho\tilde\rho$.

\subsubsection{Applications to our model}

For the density matrix $\rho_N(t)$ (\ref{normdensitymatrix}) of our model, which is
invariant under spin flip, i.e. $\tilde\rho_N = \rho_N$ and thus $\rho_N\tilde\rho_N =
\rho^{\,2}_N$, we obtain for the concurrence
\begin{eqnarray}
C\big(\rho_N(t)\big) \;&=&\; \max\big\{0,e^{-\lambda t}\big\} \;=\; e^{-\lambda t} \;,
\end{eqnarray}
and for the fully entangled fraction of $\rho_N(t)$
\begin{eqnarray}
f\big(\rho_N(t)\big) \;=\; \frac{1}{2}\big(1 + e^{-\lambda t}\big) \;,
\end{eqnarray}
which is simply the largest eigenvalue of $\rho_N(t)$. Clearly, in our case the functions
$C$ and $f$ are related by
\begin{eqnarray}
C\big(\rho_N(t)\big) \;&=&\; 2 \, f\big(\rho_N(t)\big) - 1 \;.
\end{eqnarray}
Finally, we have for the entanglement of formation of the $K^0 \bar  K^0$ system
\begin{eqnarray}\label{entanglementofformationforlambda}
E\big(\rho_N(t)\big)\;&=&\; -\frac{1+\sqrt{1-e^{-2\lambda t}}}{2}
\log_2\frac{1+\sqrt{1-e^{-2\lambda t}}}{2}\nonumber\\
& &-\frac{1-\sqrt{1-e^{-2\lambda t}}}{2}\log_2\frac{1-\sqrt{1-e^{-2\lambda t}}}{2}\;.
\end{eqnarray}

Using now our relation (\ref{zeta}) between the decoherence parameters $\lambda$ and
$\zeta$ we find a striking connection between the entanglement measure, defined by the
entropy of the state, and the decoherence of the quantum system, which describes the
amount of factorization into product states (Furry--Schr\"odinger hypothesis
\cite{Schrodinger,Furry}).\\

\textbf{Loss of entanglement:} Defining the \textit{loss of entanglement} as the gap
between an entanglement value and its maximum unity, we find
\begin{eqnarray}
1 - C\big(\rho_N(t)\big) \;&=&\; \zeta(t)\label{entanglementlossC} \;,\\
1 - E\big(\rho_N(t)\big) \;&\doteq&\; \frac{1}{\ln2}\;\zeta(t) \;\doteq\;
\frac{\lambda}{\ln2}\;t\label{entanglementlossE} \;,
\end{eqnarray}
where in Eq.(\ref{entanglementlossE}) we have expanded expression
(\ref{entanglementofformationforlambda}) for small values of the parameters $\lambda$ or
$\zeta$.\\

We get the following proposition.\\

\textbf{Proposition:}$\;$ Bertlmann--Durstberger--Hiesmayr
\cite{BertlmannDurstbergerHiesmayr2002}
\begin{itemize}
\item [$\bullet$] \textit{The entanglement loss equals the decoherence} !\\
\end{itemize}

Therefore we are able to determine experimentally the degree of entanglement of the $K^0
\bar K^0$ system, namely by considering the asymmetry (\ref{lambdaasymmetry}) and fitting
the parameter $\zeta$ or $\lambda$ to the data.

\subsubsection{Results}

\begin{figure}
\center{\includegraphics[width=250pt, height=180pt, keepaspectratio=true]{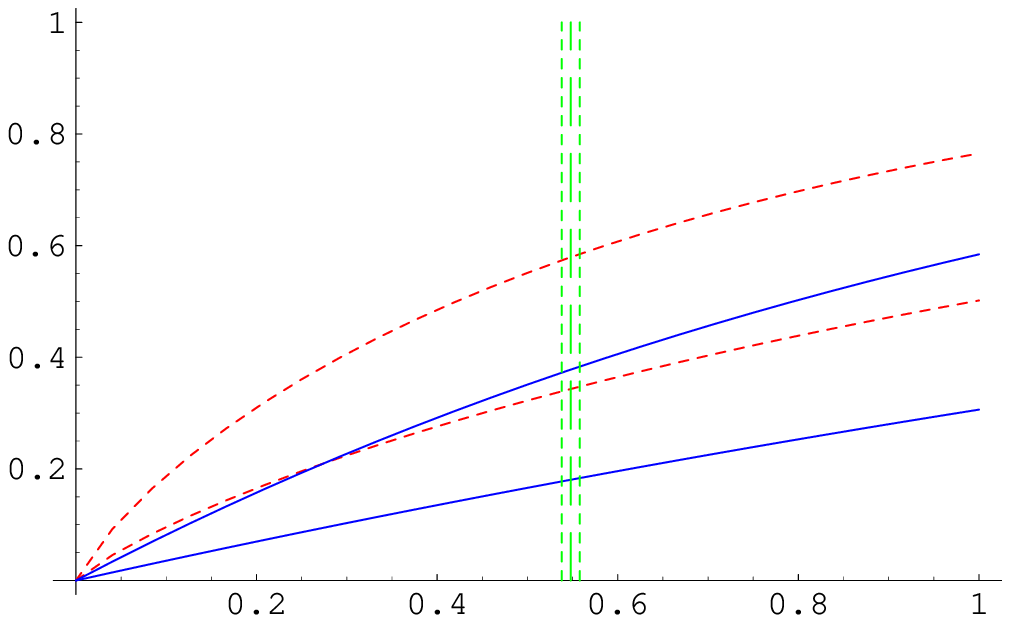}
\begin{picture}(1,1)(0,0)
\put(-7.5,4.7){absorber in the} \put(-7.5,4.3){CPLEAR experiment}
\put(-4.85,4.2){\vector(1,-1){0.75}} \put(0.05,4.2){$S(\bar\lambda_{up})$}
\put(0.05,3.3){$1-E(\bar\lambda_{up})$} \put(0.05,2.8){$S(\bar\lambda)$}
\put(0.05,1.9){$1-E(\bar\lambda)$} \put(0.2,0.1){$t/\tau_S$}
\end{picture}}
\vspace{0.5cm} \caption{The time dependence of the von Neumann entropy (dashed lines),
Eq.(\ref{vonNeumannentropy}), and the loss of entanglement of formation $1 - E$ (solid
lines), given by Eq.(\ref{entanglementofformationforlambda}), are plotted for the
experimental mean value $\bar\lambda= 1.84 \cdot 10^{-12}\;\textrm{MeV}$ (lower curve)
and the upper bound $\bar\lambda_{up} = 4.34 \cdot 10^{-12}\;\textrm{MeV}$ (upper curve),
Eq.(\ref{lambda-values}), of the decoherence parameter $\lambda$. The time $t$ is scaled
versus the lifetime $\tau_s$ of the short--lived kaon $K_S$: $t \to t/\tau_s$. The
vertical lines represent the propagation time $t_0/\tau_s \approx 0.55$ of one kaon,
including the experimental error bars, until it is measured by the absorber in the CPLEAR
experiment.}\label{entropyentanglementfigure}
\end{figure}

In Fig.\ref{entropyentanglementfigure} we have plotted the loss of entanglement $1 - E$,
given by Eq.(\ref{entanglementofformationforlambda}), as compared to the loss of
information, the von Neumann entropy function $S$, Eq.(\ref{vonNeumannentropy}), in
dependence of the time $t/\tau_s$ of the propagating $K^0 \bar  K^0$ system. The loss of
entanglement of formation increases slower with time and visualizes the resources needed
to create a given entangled state. At $t=0$ the pure Bell state $\rho^{-}$ is created and
becomes mixed for $t>0$ by the other Bell state $\rho^{+}$. In the total state the amount
of entanglement decreases until separability is achieved (exponentially fast) for $t \to
\infty$.

In case of the CPLEAR experiment, where one kaon propagates about \mbox{$2$ cm},
corresponding to a propagation time $t_0/\tau_s \approx 0.55$, until it is measured by an
absorber, the entanglement loss is about $18\%$ for the mean value and maximal $38\%$ for
the upper bound of the decoherence parameter $\lambda$.

\section{Outlook}
\label{outlook}

Entanglement is the basic ingredient for quantum communication and computation, its
effects will become important for future technologies. The fast developments in quantum
information stimulated research in particle physics, which may have effects we cannot
foresee now. We are sure that in future there will be a fruitful exchange between
particle physics and quantum optics.

There are close analogies but also significant differences between entangled
meson--antimeson systems and entangled photon or spin--$\frac{1}{2}$ systems. It turns
out that quantum mechanical tests of meson--antimeson systems are more subtle than those
of photon systems and one has to take into account the features of the mesons, which are
characteristic for such massive quantum systems. It is the decay property or a symmetry
violation property, like $CP$ violation, or the regeneration property of the quantum
state.

\subsubsection{Bell inequalities}

Generally, Bell inequalities for meson--antimeson systems contain both the freedom of
choice in time \textit{and} in quasi--spin (see Sect. \ref{kaon-Bell-CHSH}). As we
concluded in Sect. \ref{BI-time} the time variation type of BI, however, cannot be used
to exclude LRT for the familiar meson--antimeson systems due to the (unfortunate)
interplay between flavor oscillation (e.g., strangeness, beauty, charm, \ldots) and decay
time \cite{BertlmannHiesmayr2001,ghirardi91,ghirardi92,trixi}.\\

In this connection we would like to mention that the work of Ref.\cite{Go}, analyzing
entangled $B^0 \bar B^0$ meson pairs produced at the KEKB asymmetric $e^+ e^-$ collider
and collected at the BELLE detector, is hardly relevant for the test of QM versus LRT.
The reasons are twofold: Firstly, ``active'' measurements are missing, therefore one can
construct a local realistic model; secondly, the unitary time evolution of the unstable
quantum state ---the decay property of the meson--- is ignored, which is part of its
nature (for more detailed criticism, see Ref.\cite{BertlmannHiesmayrBramonGarbarino}).
Nevertheless, the work \cite{Go} represents a notable test of QM correlations exhibited
by $B^0 \bar B^0$ entangled pairs and further investigations along these lines are
recommended.\\

Considering, on the other hand, a BI at fixed time but for quasi--spin variation, a
Wigner--like BI for the several types of the K--meson (see Sect. \ref{BI-quasispin}),
provides an inequality for a symmetry violation parameter, the physical $CP$ violation
parameter (charge conjugation and parity) in our case \cite{Uchiyama,BGH-CP}.
Experimentally it is tested by studying the leptonic charge asymmetry $\delta$ of the
$K_L$ type. It is remarkable that the premises of LRT are \textit{only} compatible with
strict $CP$ conservation, i.e. with $\delta = 0$. In this way, $\delta \neq 0$ is a
manifestation of the entanglement of the considered state.
We have found the following proposition.\\

\textbf{Proposition:}$\;$ Bertlmann--Grimus--Hiesmayr \cite{BGH-CP}
\begin{itemize}
    \item [$\bullet$] \textit{$CP$ violation in $K^0 \bar K^0$ mixing leads to violation
    of a BI} !
\end{itemize}

We do believe that this connection between symmetry violation and BI violation is part of
a more general quantum feature, therefore studies of other particle symmetries in this
connection would be of high interest.\\

Bramon et al. \cite{Bramon,AncoBramon,BramonGarbarino} have established novel BI's for
entangled $K^0 \bar K^0$ pairs by using the well--known regeneration mechanism of the
kaons (see Sect. \ref{Regeneration}). After producing kaon pairs at the $\Phi$ resonance
a thin regenerator (with kaon crossing time $t_{\textrm{cross}}\ll \tau_S$) is placed
into one kaon beam near the $\Phi$ decay point. Then the entangled state contains a term
proportional to $r\, | K_L \rangle _l \otimes | K_L \rangle _r$, where $r$ denotes the
\textit{regeneration parameter}, a well--defined quantity being proportional to the
difference of the $K^0 N$ and $\bar K^0 N$ amplitudes. Allowing the entangled state to
propagate up to a time $T$, with $\tau_S \ll T \ll \tau_L$, in each beam of the two sides
either the lifetime/decay states $K_S$ versus $K_L$ or the strangeness states $K^0$
versus $\bar K^0$ are measured. Considering \textit{Clauser--Horne--type inequalities}
\cite{ClauserHorne} a set of inequalities for the parameter $R=-r\,\exp{[-i\Delta m  +
\frac{1}{2}(\Gamma_S -\Gamma_L)]\,T}$ can be derived. One of the inequalities is ---under
certain conditions (for the regeneration, etc...)--- violated by QM. It provides an
interesting experimental test at $\Phi$--factories or $p\bar p$ machines in the future.\\

\subsubsection{Decoherence}

We have developed a general but quite simple and practicable procedure to estimate
quantitatively the degree of possible decoherence of a quantum state due to some
interaction of the state with its ``environment'' (see Sect. \ref{decoherence}). This is
of special interest in case of entangled states where a single parameter, the decoherence
parameter $\lambda$ or $\zeta$ \cite{BG3,BertlmannDurstbergerHiesmayr2002,BGH},
quantifies the strength of decoherence between the two subsystems or the amount of
spontaneous factorization of the wavefunction. The asymmetry of like-- and unlike--flavor
events is directly sensitive to $\lambda$ or $\zeta$.

On the other hand, we have found a bridge to common entanglement measures, such as
entanglement of formation and concurrence (see Sect. \ref{Entanglement-Measure}), and
could derive the Proposition that \textit{the entanglement loss equals the
decoherence}$\,$! In this way we also measure the amount of entanglement of the state.

Using experimental data of the CPLEAR experiment provides us with upper bounds to the
decoherence and the entanglement loss of the $K^0 \bar K^0$ state produced in $p \bar p$
collisions, which is of macroscopic extent ($9$ cm).

As up to now only two data points are available, further experimental data are certainly
highly desirable and would sharpen the bounds considerably. Such data could be collected
at the $\Phi$--factory DA$\Phi$NE in Frascati. In this connection, a recent analysis of
entangled $K_S K_L$ systems by the KLOE collaboration \cite{KLOE-Domenico} sets new
bounds on the decoherence parameter $\zeta$.

Indeed, it would be of great interest to measure in future experiments the asymmetry of
like-- and unlike--flavor events for several different times, in order to confirm the
time dependence of the decoherence effect. In fact, such a possibility is now offered in
the $B$-meson system. As already mentioned (Sect. \ref{Measurement}) entangled $B^0\bar
B^0$ pairs are created with high density at the asymmetric $B$-factories and identified
by the detectors BELLE at KEK-B \cite{Belle,Leder} and BABAR at PEP-II
\cite{Aubert,Babar} with a high resolution at different distances or times. On the basis
of the time dependent event rates we could test experimentally the predictions of our
decoherence model.

Of course, these decoherence investigations could be performed with other entangled
systems, like photonic systems, as well.

\subsubsection{Complementarity}

Meson systems in particle physics are also an interesting testing ground for basic
principles of QM such as \textit{quantitative wave--particle duality} \cite{SBGH3,SBGH4}
or \textit{quantum marking} and \textit{quantum erasure} \cite{SBGH1,SBGH2}.

Bohr's \textit{complementarity principle} (or \textit{duality principle}) can be
formulated in a quantitative way \cite{GreenbergerYasin,Englert} by defining on one hand
the \textit{fringe visibility} ---the sharpness of the interference pattern, the
wave--like property--- and on the other the \textit{path predictability}, e.g., in a
two--path interferometer \cite{RauchWerner}. This ``which--way'' knowledge on the path
taken by the interfering system expresses the particle--like property of the system.

Considering now a produced $K^0$ beam, it oscillates between $K^0$ and $\bar K^0$ but it
also represents a superposition of $K_S$ and $K_L$ which decay at totally different rates
(recall Sect. \ref{QM-K-meson}). Then the strangeness oscillations can be viewed as
fringe visibility and the $K_S$ and $K_L$ states with their different decay distances as
the analogues of the two paths in the interferometer. The comparison of present data with
the calculations of this kind of \textit{kaon interferometry} satisfies nicely the
statement of \textit{quantitative duality} \cite{SBGH3,SBGH4}. Further experiments, e.g.,
at the $\Phi$--factory DA$\Phi$NE are of interest in this connection.

\subsubsection{Geometric phase --- entanglement}

An interesting question is: how can we influence the entanglement of a
spin--$\frac{1}{2}$ system by creating a geometric phase? Geometric phases such as the
Berry phase \cite{Berry1} play a considerable role in physics and arise in a quantum
system when its time evolution is cyclic and adiabatic. There is increasing interest in
combining both the geometric phase and the entanglement of a system
\cite{Sjoqvist2000,TongKwekOh2003,MilmanMosseri2003,TongSjoqvistKwekOhEricsson,BDHH}.

In Ref.\cite{BDHH} we have studied the influence of the Berry phase on the entanglement
of a spin--$\frac{1}{2}$ system by generating the Berry phase with an adiabatically
rotating magnetic field in one of the paths of the particles, and we have eliminated the
dynamical phase ---being sensitive just to the geometric phase--- by a ``spin--echo''
method. We have considered a \textit{pure} entangled system where a phase
---like our geometrical one--- does \textit{not} change the amount of entanglement and
therefore \textit{not} the extent of nonlocality of the system, which is determined by
the maximal violation of a BI. In our case the Berry phase just affects the Bell angles
in a specific way. On the other hand, keeping the measurement planes fixed the polar Bell
angles vary and the maximum of the $S$-function of the CHSH inequality (recall Sect.
\ref{Bell-inequalities}) varies with respect to the Berry phase.

We have applied the investigation to neutron interferometry where one can achieve
entanglement between different degrees of freedom ---an internal and external one--- the
spin and the path of the neutron. In this case it is physically rather contextuality than
nonlocality which is tested experimentally
\cite{BDHH,RauchWerner,BasuBandyopadhyayKarHome,HasegawaLoidlBadurekBaronRauch}.

There are still open problems in how does entanglement behave ---the several entanglement
measures--- if geometric phases are introduced into \textit{mixed} quantum states. How do
they change entanglement? What happens to the Berry phase of the system if in addition we
allow for interaction of the quantum system with an environment, thus allowing for
decoherence? How can it be experimentally realized? Work along these lines is of
increasing interest
\cite{FonsecaAguiarThomaz,KamleitnerCresserSanders,Carollo-Vedral,DoddHalliwell,WhitneyGefen}.

\subsubsection{Generalized Bell inequality --- entanglement witness}

In quantum communication, the two remote parties, Alice and Bob, want to carry out a
certain task with minimal communication. Comparing now the amount of communication
necessary by using classical \textit{or} quantum bits, it turns out that sharing
entangled quantum states ---instead of classical or separable ones--- leads to savings in
the communication. That makes entangled states so important in quantum communication
technology.

But how do we witness an entangled state? Via the familiar Bell inequalities described
before (Sect. \ref{Bell-inequalities})? \textit{Yes}, if the states are \textit{pure}
entangled states but \textit{no} if they are \textit{mixed} entangled. Werner
\cite{Werner} found out that a certain mixture of entangled and separable states,
so--called \textit{Werner states}, do satisfy the familiar BI. Thus the importance of the
BI is that they serve as criterion for nonlocality ---as departures from LRT--- and they
are valuable when considering phenomena like quantum teleportation or quantum
cryptography. But as a criterion for separability they are rather poor.\\

There exist, however, some operators, so--called \textit{entanglement witnesses},
satisfying an inequality, a \textit{generalized Bell inequality} (GBI), for \textit{all}
separable states and violating this inequality for an entangled state
\cite{BNT,HHH96,TerhalPhysLett,Bruss}.

In Ref.\cite{BNT} we have studied the class of entanglement witnesses; considering
density matrices (states) and operators (entanglement witnesses) as elements of a Hilbert
space it turns out that entanglement witnesses are tangent functionals to the set of
separable states. Considering the Euclidean distance of the vectors in Hilbert space for
entangled and separable states we have found the following Theorem.\\

\textbf{Theorem:}$\;$ Bertlmann--Narnhofer--Thirring \cite{BNT}
\begin{itemize}
    \item [$\bullet$] \textit{The Euclidean distance of an entangled state to the
    separable states is equal to the maximal violation of the GBI with the tangent
    functional as entanglement witness} !\\
\end{itemize}

Viewing the Euclidean distance as entanglement measure we have found a nice geometric
picture for entanglement (and its value), for GBI (and its maximal violation) and for the
tangent functional which characterizes (as a tangent) the set of separable states. All
three conceptions are only different aspects of the same geometric picture. It is
especially illustrative for the example of two spins, Alice and Bob. In this connection
investigations for multi--particle entangled states, e.g., GHZ--states
\cite{GHZ,GHZ-Shimony}, are certainly of high interest.

\vspace{1cm}

\section{Acknowledgement}

The author wants to take the opportunity to thank his collaborators and friends for all
the joyful discussions about this field, special thanks go to Markus Arndt, Katharina
Durstberger, Gerhard Ecker, Stefan Filipp, Walter Grimus, Yuji Hasegawa, Beatrix
Hiesmayr, Heide Narnhofer, Helmut Neufeld, Herbert Pietschmann, Helmut Rauch, Walter
Thirring and Anton Zeilinger. The aid of EU project EURIDICE EEC-TMR program
HPRN-CT-2002-00311 is acknowledged. Finally, I would like to thank the organizers of the
Schladming School for creating such a stimulating scientific atmosphere.


%
%
%

%
%



\printindex
\end{document}